\documentclass[fleqn,usenatbib]{mnras}
\usepackage{newtxtext,newtxmath}

\usepackage[T1]{fontenc}

\DeclareRobustCommand{\VAN}[3]{#2}
\let\VANthebibliography\thebibliography
\def\thebibliography{\DeclareRobustCommand{\VAN}[3]{##3}\VANthebibliography}

\usepackage{graphicx}	
\usepackage{amsmath}	
\usepackage{mathrsfs} 
\usepackage{stfloats}
\usepackage{orcidlink}

\newcommand{\angstrom}{\mbox{\normalfont\AA}}

\newcommand{\new}[1]{\textcolor{red}{#1}}

\newcommand{\Lya}{Ly$\alpha$}
\newcommand{\HI}{H\,\textsc{I}}
\newcommand{\HeII}{He\,\textsc{II}}
\newcommand{\SiII}{Si\,\textsc{II}}
\newcommand{\SiIII}{Si\,\textsc{III}}
\newcommand{\SiIV}{Si\,\textsc{IV}}
\newcommand{\CIV}{C\,\textsc{IV}}
\newcommand{\MgII}{Mg\,\textsc{II}}

\title[An improved model for \SiIII\ absorption]{An improved model for the effect of correlated  \SiIII\ absorption on the one-dimensional Lyman-$\alpha$ forest power spectrum}

\author[Ke Ma et al.]{Ke Ma$^{1}\,\orcidlink{0000-0002-0564-891X}$\thanks{E-mail: ke.ma@nottingham.ac.uk}, James S. Bolton$^{1}\,\orcidlink{0000-0003-2764-8248}$, Vid Ir\v{s}i\v{c}$^{2}\,\orcidlink{0000-0002-5445-461X}$, Prakash Gaikwad$^{3}\,\orcidlink{0000-0002-2423-7905}$ and Ewald Puchwein$^{4}\,\orcidlink{0000-0001-8778-7587}$
\\
$^{1}$School of Physics and Astronomy, The University of Nottingham, University Park, Nottingham, NG7 2RD, UK\\
$^{2}$Centre for Astrophysics Research, Department of Physics, Astronomy and Mathematics, University of Hertfordshire, College Lane, Hatfield, AL10 9AB, UK\\
$^{3}$Department of Astronomy, Astrophysics and Space Engineering, Indian Institute of Technology Indore, Simrol, MP 453552, India\\
$^{4}$Leibniz-Institut f\"ur Astrophysik Potsdam, An der Sternwarte 16, 14482 Potsdam, Germany\\
}

\date{Accepted XXX. Received YYY; in original form ZZZ}

\pubyear{2025}

\begin{document}
\maketitle

\begin{abstract}
We present an analysis of \SiIII\ absorption and its effect on the 1D \Lya\ forest power spectrum using the Sherwood-Relics hydrodynamical simulation suite.  In addition to oscillations from the \Lya--\SiIII\ cross correlation that are damped toward smaller scales, we find an enhancement in small-scale power that has been ignored in previous studies.  We therefore develop a new analytical fitting function that captures two critical effects that have previously been neglected: distinct Ly$\alpha$ and \SiIII\ line profiles, and a variable ratio for coeval Ly$\alpha$ and \SiIII\ optical depths. In contrast to earlier work, we also predict amplitudes for the \SiIII\ power spectrum and \Lya-\SiIII\ cross power spectrum that decrease toward lower redshift due to the hardening metagalactic UV background spectrum at $z\lesssim 3.5$.  The fitting function is validated by comparison against multiple simulated datasets at redshifts $2.2\leq z \leq 5.0$ and wavenumbers $k < 0.2\rm\,s\,km^{-1}$.  Our model has little effect on existing warm dark matter constraints from the \Lya\ forest when adopting a physically motivated prior on the silicon abundance. It will, however, be an essential consideration for future, high precision \Lya\ forest power spectrum measurements.

\end{abstract}



\section{Introduction}
\label{sec:intro}

The \Lya\ forest, characterised by absorption features imprinted on the spectra of distant quasars, is produced when light from background quasars is absorbed by neutral hydrogen in the intergalactic medium (IGM). It stands as one of the most informative probes for studying the structure and evolution of the Universe \citep{Meiksin2009, Mcquinn2016}. As the light traverses regions of varying hydrogen gas density along the line of sight, a complex pattern of absorption emerges. This pattern reflects both the physical state of the IGM and the underlying dark matter distribution, making the \Lya\ forest an essential tool for investigating a range of astrophysical and cosmological phenomena.

One of the most widely used summary statistics used for extracting information from the Ly$\alpha$ forest is the one-dimensional Ly$\alpha$ power spectrum, which quantifies the statistical distribution of absorption features as a function of spatial scale. Its study remains an important area of research. In particular, it plays a crucial role in constraining cosmological parameters \citep[e.g.,][]{Croft2002,Viel2004,Seljak2005,Mcdonald2006}, dark matter models \citep[e.g.,][]{Narayanan2000,Viel2013,Rogers2021,Hooper2022,Villasenor2023,Irsic2024,GarciaGallego2025} and primordial magnetic fields \citep{Pavicevic2025}, complementing observations from the cosmic microwave background, galaxy clustering, and weak gravitational lensing.

Current measurements of the 1D power spectrum are divided into two regimes. High-resolution spectroscopic surveys (e.g., Keck/HIRES, VLT/UVES; $R \gtrsim 20,000$, S/N $> 10$) resolve small-scale features but have limited sample sizes ($\lesssim 100$ QSOs), leading to large statistical uncertainties \citep[e.g.,][]{Irsic2017b, Boera2019, Karacayli2022}. Representing the first regime, \cite{Boera2019} measured the power spectrum at $z =$ 4.2, 4.6, and 5.0 using fifteen high-resolution QSO spectra from Keck/HIRES and VLT/UVES, reaching scales down to  $k \sim 0.2$ s km$^{-1}$ with $\sim 15$ per cent precision. Conversely, surveys like SDSS/eBOSS and DESI achieve larger samples ($\sim 10^{5}$ QSO spectra) at the cost of spectral resolution, limiting their sensitivity to wavenumbers beyond $k \sim 0.03 \rm\,s\,km^{-1}$. For the second regime, the most recent public results come from SDSS/eBOSS \citep{Chabanier2019}, measuring the power spectrum from $z=2.2$ to 4.6 and extending to $k \sim 0.02$ s km$^{-1}$, and DESI DR1 \citep{Ravoux2025, Karacayli2025}, measuring from $z=2.2$ to 4.2 and extending to $k \sim 0.026 - 0.039$ s km$^{-1}$ depending on redshift.

Accurate modeling of the Ly$\alpha$ forest power spectrum, however, faces challenges due to contamination from metal absorption lines. These lines are broadly categorised as either correlated or uncorrelated with  Ly$\alpha$ absorption. Correlated metal lines trace the Ly$\alpha$ forest due to their spectral and redshift proximity within the same large-scale structures. The most significant contributor is \SiIII\ ($\lambda$1207), arising from doubly ionised silicon. Uncorrelated metal lines, conversely, do not align spatially or in redshift with the Ly$\alpha$ forest and often trace distinct astrophysical environments, resulting in a stochastic relationship with Ly$\alpha$ absorption. Prominent examples include \CIV\ ($\lambda \lambda$1548, 1551), \SiIV\ ($\lambda \lambda$1394, 1403), and \MgII\ ($\lambda \lambda$2796, 2804). These metal lines introduce systematic distortions that complicate cosmological measurements and IGM modelling using the \Lya\ forest power spectrum. Consequently, improving models of their impact on the \Lya\ forest power spectrum is essential.

In practice, the effect of uncorrelated metals on the 1D power spectrum can be reduced to negligible levels by estimating their power spectrum using sidebands -- spectral regions devoid of Ly$\alpha$ absorption that lie redward of the QSO \Lya\ emission lines \citep[e.g.,][]{Palanque-Delabrouille2013, Chabanier2019, Ravoux2023, Boera2019, Karacayli2022, Karacayli2024}. Contamination from correlated metals, however, is harder to isolate due to the fact they are blended in the Ly$\alpha$ forest. The canonical modelling approach, first introduced by \cite[][hereafter M06]{Mcdonald2006}, assumes that the transmitted flux contrast of \SiIII\ absorption can be obtained by a linear rescaling of the Ly$\alpha$ flux contrast. This ansatz yields an expression for the \SiIII\ power spectrum in terms of the Ly$\alpha$ power spectrum. At large scales, it is well-established that this model accurately describes the oscillatory behavior in the power spectrum induced by \SiIII\ absorption.  Furthermore, the resulting systematic errors from the \SiIII\ oscillations are small compared to other uncertainties. 

However, the M06 ansatz will break down towards smaller scales as differences in individual absorption line profiles become significant.  Furthermore, the redshift evolution of the \SiIII\ power spectrum is also uncertain.  These limitations pose problems for high-resolution measurements of the power spectrum, particularly as statistical errors become smaller.  To address this issue, this paper presents a revised approach for the impact of correlated \SiIII\ absorption on the Ly$\alpha$ forest spectrum.  We develop a physically motivated fitting function based on the Sherwood-Relics simulation suite \citep{Bolton2017, Puchwein2023}, from which we directly compute Ly$\alpha$ and \SiIII\ optical depths and construct mock QSO spectra.  

The paper is organised as follows: Section~\ref{sec:method} outlines our methodology for creating the mock QSO spectra and quantifying the power spectrum. Section~\ref{sec:modelling} discusses the limitations of previous work and an analytical argument that helps build insight into our revised model. We test and validate our model against simulation datasets with varying redshifts, temperatures, and metallicities in Section~\ref{sec:validation}.  Readers who are primarily interested in how to implement our model within parameter inference frameworks may wish to skip directly to Section~\ref{sec:usage} and Eq.~(\ref{eq:finalfit}).  The implications of our model for warm dark matter constraints are discussed briefly in Section~\ref{sec:comparedata}. Finally, Section~\ref{sec:summary} summarises our results and conclusions.  


\section{Numerical method} \label{sec:method}
\subsection{Hydrodynamical simulation} \label{sec:simulation}

\begin{figure*}
    \centering
        \includegraphics[width=0.95\textwidth]{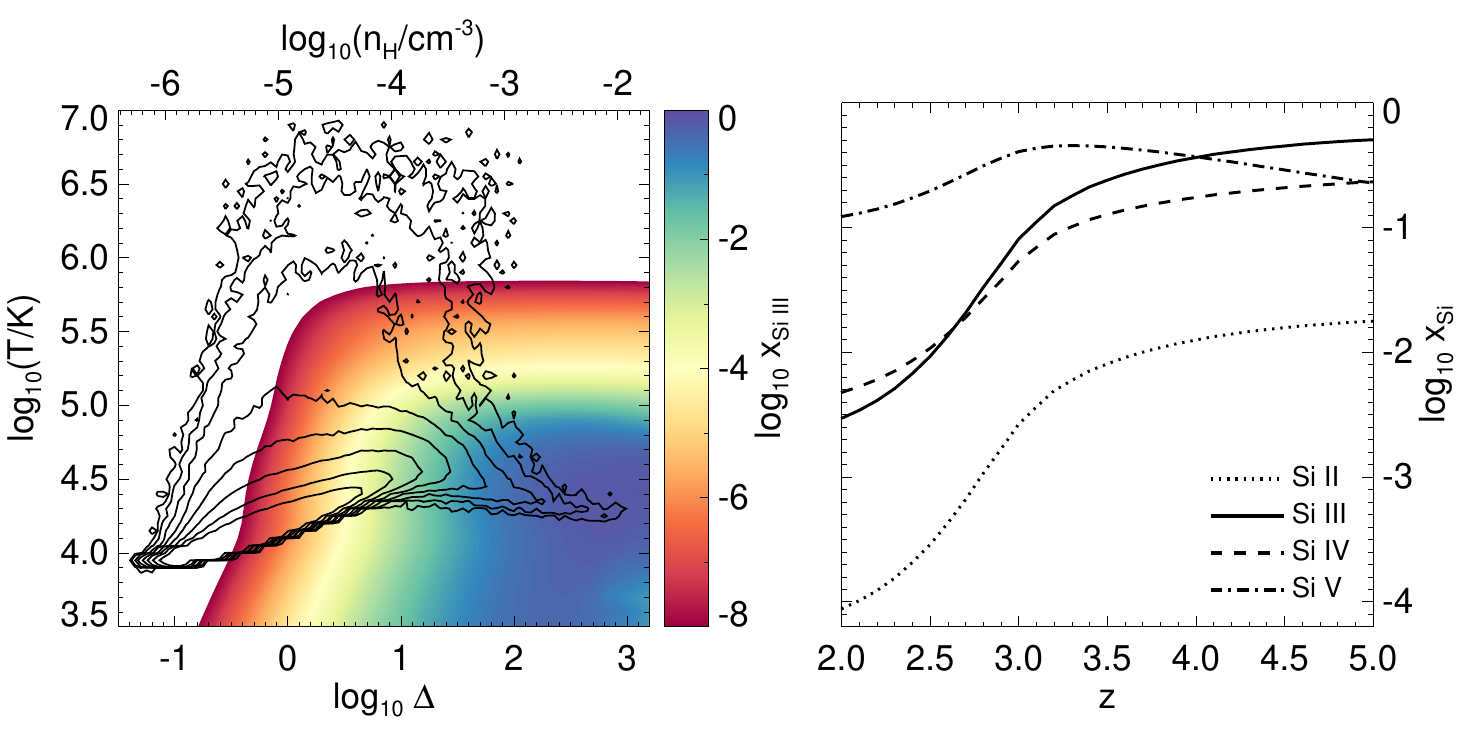}
    \vspace{-0.4cm}
    \caption{Left panel: The solid black contours show the volume weighted temperature-density plane at $z=3$ for the Sherwood-Relics hydrodynamical simulation used in this work.   The number density of points increases by $0.5$ dex within each contour level.  Note the horizontal axis is given in terms of both the logarithm of the gas overdensity, $\log_{10}\Delta$, and the logarithm of the hydrogen number density, $\log_{10}(n_{\rm H}/\rm cm^{-3})$.  The overlaid colour scale shows the logarithm of the \SiIII\ fraction, $\log_{10}x_{\rm SiIII}$, predicted by the photo-ionisation code Cloudy \citep{Ferland1998,Chatzikos2023} using the UV background synthesis model from \citet{Puchwein2019}. Right panel:  The redshift evolution of silicon ion fractions for gas with density $\log_{10}(n_{\rm H}/\rm cm^{-3})=-3.4$ and temperature $\log_{10}(T/\rm K)=4.5$. The decline in the \SiII, \SiIII\ and \SiIV\ fractions at $z\lesssim 3.5$ is driven by the hardening of the UV background spectrum due to ionising photons emitted by quasars.}
    \label{fig:ionfrac}
\end{figure*}

We use a simulation from the Sherwood-Relics project \citep{Puchwein2023} to construct \Lya\ ($\lambda1216$) and \SiIII\ ($\lambda1207$) absorption spectra.  Sherwood-Relics is an extensive series of high-resolution cosmological hydrodynamical simulations performed using a customised version of \textsc{P-Gadget-3} \citep{Springel2005}.  In this work we use a cosmological box of size $40h^{-1}\rm\, cMpc$ with $2 \times 2048^{3}$ dark matter and gas particles. The dark matter particle mass is $M_{\rm dm}=5.37\times 10^{5}h^{-1}\,\rm M_{\odot}$ and the gas particle mass is $M_{\rm gas}=9.97\times 10^{4}h^{-1}\,M_{\odot}$.  The box size and mass resolution have been chosen to resolve the small-scale structure of the \Lya\ forest \citep[see e.g.][]{BoltonBecker2009}.  

The initial conditions for the simulation are identical to those used in the earlier Sherwood simulation project \citep{Bolton2017}.   We assume a flat $\Lambda$CDM cosmology with $\Omega_{\Lambda} = 0.692$, $\Omega_{\rm M} = 0.308$, $\Omega_{\rm b}=0.0482$, $\sigma_{8} = 0.829$, $n_{\rm s} = 0.961$, $h = 0.678$, and a primordial helium mass abundance of $Y = 0.24$.  A computationally efficient star-formation scheme is used, where gas particles are converted into collisionless star particles if they have gas overdensities $\Delta > 10^{3}$ and gas temperatures $T < 10^{5}\rm\, K$ \citep{Viel2004}.   The simulation uses a spatially uniform ultraviolet (UV) background \citep{Puchwein2019} and the thermal and ionisation state of the hydrogen and helium gas is computed using a non-equilibrium thermo-chemistry solver \citep{Puchwein2015}.

Periodic sightlines are drawn through the simulation to extract the gas densities, temperatures, peculiar velocities and \HI\ fractions needed to compute mock \Lya\ absorption spectra.  We use the interpolation scheme described by \citet{Theuns1998} for this purpose.  The sightlines are extracted at a redshift interval of $\Delta z =0.1$, and for each redshift we draw a total of $5000$ such sightlines with $2048$ pixels, giving a pixel size of $19.5h^{-1}\rm\,ckpc$.  The \Lya\ optical depths along each sightline are then computed by performing a convolution with the the Voigt line profile from \citet{TepperGarcia2006}.   

We use a similar approach to construct the \SiIII\ absorption spectra, although with some additional steps.    Firstly, the Sherwood-Relics simulations do not use a sub-grid model for metal production and dispersal.  Instead, we assume a fiducial silicon abundance model that follows the gas overdensity, $\Delta=\rho/\langle \rho \rangle$, such that 
\begin{equation} [{\rm Si/H}] =  - 2.70 + 0.08(z-3) +0.65(\log_{10}\Delta -0.5). \label{eq:SiH} \end{equation}
\noindent
Eq.~(\ref{eq:SiH}) follows from the pixel optical depth silicon abundance constraints presented by \citet{Schaye2003} and \citet{Aguirre2004} using data at $z=1.8$--$4.1$.\footnote{In more detail,  we use $[{\rm Si/H}]=[{\rm C/H}] + [{\rm Si/C}]$, with $[{\rm C/H}]$ from eq. (8) in \citet{Schaye2003} and $[{\rm Si/C}]=0.77$ from \citet{Aguirre2004}.  These metallicity constraints were derived using ionisation corrections based on the \citet{HaardtMadau2001} UV background model.}   The \SiIII\ optical depth, $\tau_{\rm \SiIII}$, is then proportional to
\begin{equation} \tau_{\rm \SiIII} \propto x_{\rm \SiIII}(\Delta,T)10^{[{\rm Si/H}] + ({\rm Si/H})_{\odot}}, \label{eq:tau_SiH} \end{equation}
\noindent
where $x_{\rm \SiIII}$ is the \SiIII\ fraction (which is, in general, density and temperature dependent given some UV background model) and $({\rm Si/H})_{\odot}=-4.49$ is the solar abundance for silicon \citep{Asplund2009}.  We note, however, that the precise form of the silicon distribution is not critical for our analysis.  We have verified that adopting a constant silicon abundance instead of Eq.~(\ref{eq:SiH}) yields a qualitatively similar effect on the shape of the (\SiIII\ contaminated) \Lya\ forest power spectrum (see also Appendix~\ref{app:zscale_test}).

The equilibrium ionisation balance of the silicon is calculated using the photo-ionisation code Cloudy \citep{Ferland1998,Chatzikos2023}.  We use Cloudy in combination with the \citet{Puchwein2019} UV background spectrum to create a three-dimensional look-up table listing the the \SiIII\ fraction, $x_{\rm \SiIII}$, as a function of the logarithm of the gas temperature, $\log_{10}(T/\rm K)$, the logarithm of the hydrogen number density, $\log_{10}(n_{\rm H}/\rm cm^{-3})$ and redshift, $z$.  The table data are linearly interpolated to obtain the $x_{\rm \SiIII}$ for each pixel in our simulated sightlines.   

An example of the predicted  \SiIII\ fraction as a function of temperature and density are shown in the left panel of Figure~\ref{fig:ionfrac}.  The black contours display the volume-weighted temperature-density plane for gas in the hydrodynamical simulation at $z=3$, with a colour scale displaying the logarithm of the \SiIII\ fraction.  The right panel of Figure~\ref{fig:ionfrac} shows the redshift evolution of various silicon ion fractions for gas with a density and temperature typical of the \SiIII\ absorption, $\log_{10}(n_{\rm H}/\rm cm^{-3})=-3.4$ and $\log_{10}(T/\rm K)=4.5$.  The \SiIII\ fraction is largest at $z=5$, and the decline in the \SiII, \SiIII\ and \SiIV\ fractions at $z\lesssim 3.5$ is driven by the hardening of the UV background spectrum due to ionising photons emitted by quasars (i.e. the same spectral hardening that drives \HeII\ reionisation, e.g. \citet{Gaikwad2025}).  Note also that for the UV background model and redshift range considered in this study, \SiIII\ (with ionisation potential $33.5\rm\,eV$) is more abundant than \SiII\ (with ionisation potential $16.3\rm\,eV$), where the latter will also correlate with \Lya\ for the transitions \SiII\ ($\lambda \lambda$1190, 1193) and \SiII\ ($\lambda$1260).  We initially included these \SiII\ absorption lines in our analysis, but neglected them after establishing their contribution to the 1D power remains small compared to \SiIII\ in our model.  We note, however, that if the \SiII\ absorption is typically associated with dense circumgalactic gas that self-shields to hydrogen ionising photons, it will be underestimated in our model due to the lack of self-shielded gas with $\Delta>10^{3}$ in Sherwood-Relics.

Finally, before performing the Gaussian line profile convolution to obtain the \SiIII\ optical depths, we resample the sightline quantities onto a finer grid with pixel size $9.8h^{-1}\rm\,ckpc$. This resampling is needed to resolve the thermal Doppler widths of the silicon absorbers, $b_{\rm \SiIII}=2.4{\rm\,km\,s^{-1}}(T/10^{4}\rm\,K)^{1/2}$, and thus ensure the \SiIII\ optical depths are converged. Note also that the Ly$\alpha$ and \SiIII\ optical depths are all extracted from the same randomly selected sightlines, ensuring the absorption traces the same physical structures in the simulation.

\subsection{Mock quasar sightlines}

\begin{figure*}
    \centering
    \includegraphics[width=2\columnwidth]{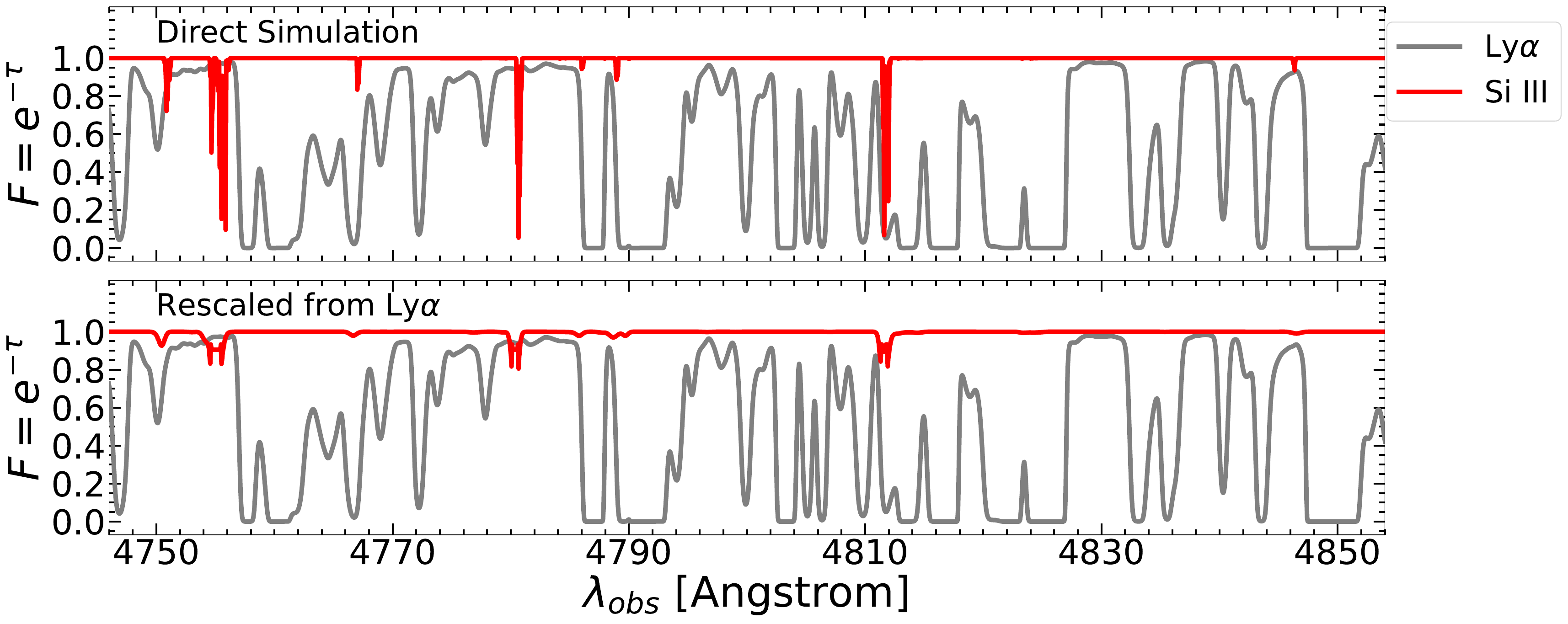}
    \vspace{-0.3cm}
    \caption{The transmitted flux for the Ly$\alpha$ forest (grey) and \SiIII\ absorption (red) for a random simulated sightline at $z \sim 3$. The upper panel shows the results from our direct calculation of the \SiIII\ spectrum from the hydrodynamical simulation described in Section~\ref{sec:method}. The lower panel compares this to the \SiIII\ absorption obtained by a rescaling of the \Lya\ optical depths.  This is similar to the approach often used in the literature when modelling the contribution of correlated metal absorption to the 1D \Lya\ forest power spectrum.   In this example the scaling is chosen so that the mean \SiIII\ transmission in both panels is equal.  Note the narrower widths for the \SiIII\ absorption in the direct calculation.  The Ly$\alpha$ absorption is identical in both panels.
    }
    \label{fig:Flux_Si3compare}
\end{figure*}

Once the optical depths are extracted from the hydrodynamical simulations, we combine the individual sightlines drawn from the hydrodynamical simulation into mock quasar spectra.  First, for a given quasar redshift $z_{\rm QSO}$, we construct a full-length \Lya\ forest spectrum (i.e. from $\lambda_{\rm \alpha}(1+z_{\rm QSO})$ to $\lambda_{\rm \beta}(1+z_{\rm QSO})$, where $\lambda_{\rm \alpha}$ and $\lambda_{\rm \beta}$ are the \Lya\ and Ly$\beta$ rest frame wavelengths) by combining the simulated sightlines at the required redshifts.  These are stitched together so that the optical depth remains continuous; the periodic nature of the sightlines allows us to shift each segment to maintain continuous intersections.  The same process is used to extract the coeval \SiIII\ absorption spectra over a matching length scale.  The Ly$\alpha$ absorption is furthermore normalised to match the effective optical depth measured by \cite{Becker2013} at the appropriate redshifts.  Finally, the \SiIII\ optical depths are shifted to align at the correct observed wavelengths and then added to the \Lya\ optical depths, where we assume the rest frame wavelengths of the transitions are Ly$\alpha$ ($\lambda$1216) and \SiIII\ ($\lambda$1207).   An example of the predicted absorption in a noiseless spectrum at $z\sim 3$ is displayed in the upper panel of Figure~\ref{fig:Flux_Si3compare}.

\subsection{Computing the power spectrum}

We shall use the noiseless mock spectra to compute the 1D \Lya\ forest power spectrum, both including and excluding the correlated \SiIII\ absorption.  Following the approach commonly adopted in the literature \citep[e.g.][]{Palanque-Delabrouille2013, Chabanier2019, Ravoux2023, Karacayli2024}, we compute the power spectrum, $P(k)$, using the Fast Fourier Transform (FFT) approach, where
\begin{equation}
    P(k) \propto | \tilde{\delta}(k)|^2.
    \label{eqn:FFT}
\end{equation}
Here $\tilde{\delta}(k)$ is the Fourier transform of the flux contrast, $\delta(v)$, which is given by:
\begin{equation}
    \delta(v) = \frac{F(v)}{\overline{F}(v)} - 1,
    \label{eqn:deltaF_Lya}
\end{equation}
where $F(v)$ is the transmitted flux in a pixel at velocity $v$, and $\overline{F}(v)$ is the mean transmission of the intervening \Lya\ forest. For the latter we assume the effective optical depth, $\tau_{\rm eff}$, reported by \cite{Becker2013}, where $\overline{F}(\new{v})=e^{-\tau_{\rm eff}}$. 


\section{Modelling the \Lya-\SiIII\ correlation} \label{sec:modelling}

\subsection{The M06 analytical model}

\begin{figure*}
    \centering
    \includegraphics[width=2\columnwidth]{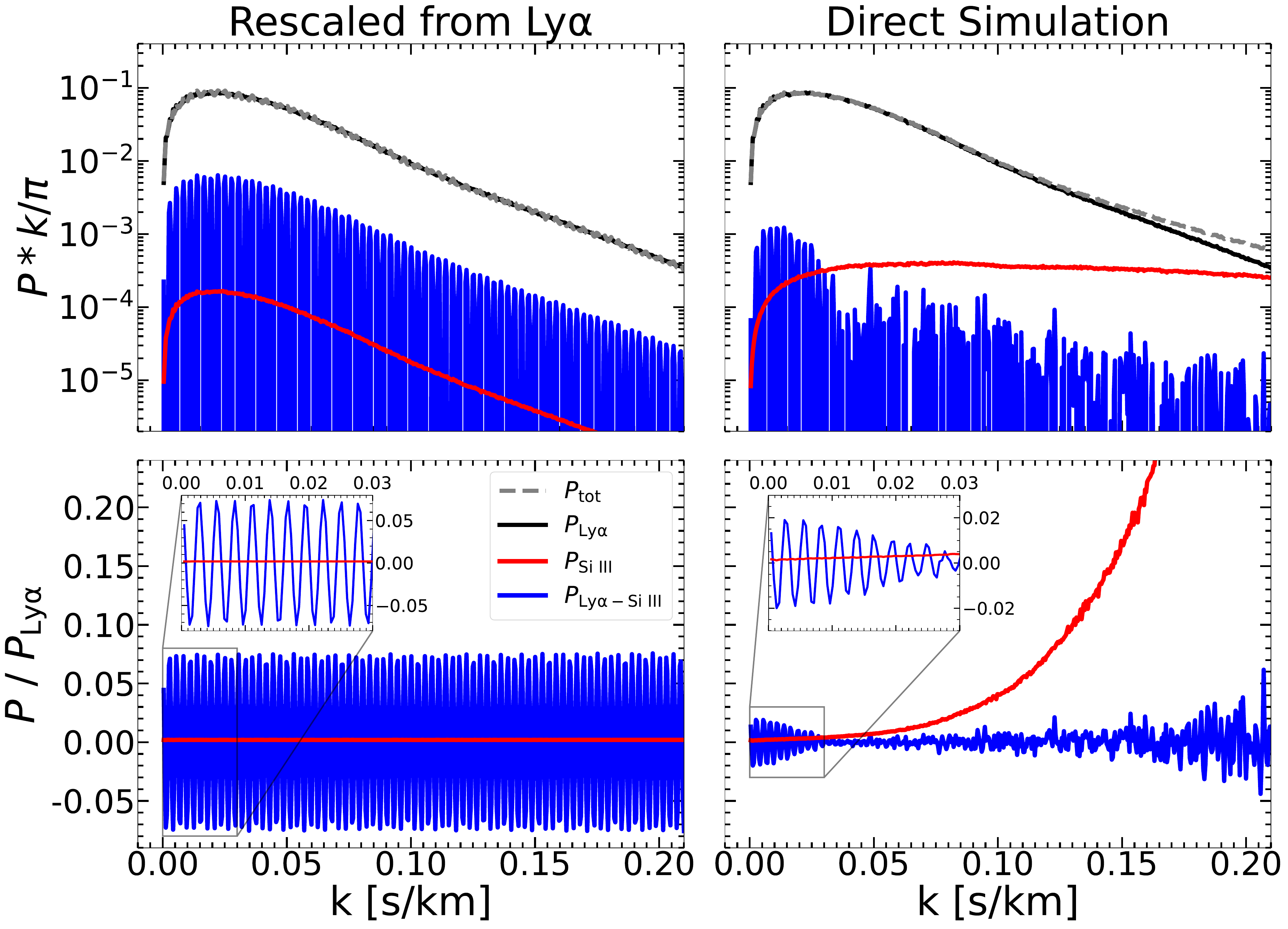}
    \vspace{-0.3cm}
    \caption{A comparison of the total \Lya\ forest power spectrum including the contamination by correlated \SiIII\ absorption ($P_{\rm tot}(k)$, grey dashed curves) and its constituent terms (see Eq.~\ref{eqn:Pclm_old}) at $z=3$, where the \SiIII\ absorption has been computed either by a linear rescaling of the \Lya\ optical depths (left column) or from a direct simulation following the procedure described in Section~\ref{sec:method} (right column).   Additional curves show the power spectrum for Ly$\alpha$ absorption only ($P_{\rm Ly\alpha}(k)$, black solid curve), \SiIII\ absorption only ($P_{\rm \SiIII}(k)$, red solid curve), and the cross term between Ly$\alpha$ and \SiIII\ ($P_{\rm Ly\alpha-\SiIII}(k)$, blue solid curve). The bottom panels show the ratio of $P_{\text{Si-III}}(k)$ and $P_{\rm Ly\alpha-\SiIII}(k)$ to $P_{\rm Ly\alpha}(k)$.  In the left column, the value for the linear rescaling parameter $A$, where $\tau_{\rm \SiIII}(v)=A\tau_{\rm Ly\alpha}(v+\Delta v)$, was chosen so that the amplitude of $P_{\rm \SiIII}(k)$ is similar to the direct simulation on large scales.}  
    \label{fig:Pclm_compare}
\end{figure*}

It is well known that the correlations between metal line transitions with wavelengths similar to the \Lya\ transition are an important systematic in analyses of the 1D \Lya\ forest power spectrum.  For \SiIII, which is one of the most prominent correlated metals, a feature at $\Delta v=2271\rm\,km\,s^{-1}$ is produced in the \Lya\ forest auto-correlation.  This also manifests as an oscillatory feature in the 1D power spectrum with peak separations of $\Delta k = 2\pi / \Delta v$.  

Hence, in order to obtain a good match between observations and models, the contribution of \SiIII\ to the power spectrum is usually treated as a nuisance parameter.  M06 first outlined the model framework adopted in the literature.   These authors made the ansatz that the \SiIII\ transmitted flux contrast, $\delta_{\rm \SiIII}$, will have the same shape as the \Lya\ forest transmitted flux contrast, $\delta_{\rm Ly\alpha}$, up to some normalisation factor $a$, where for a pixel at velocity $v$,
\begin{equation}
\begin{aligned}
    \delta_{\rm tot}(v) &= \delta_{\rm Ly\alpha}(v) + \delta_{\rm \SiIII}(v), \\
    &= \delta_{\rm Ly\alpha}(v) + a\ \delta_{\rm Ly\alpha}(v + \Delta v).
\end{aligned}
\label{eqn:dclm_old}
\end{equation}
\noindent
Here $\Delta v=2271\rm\,km\,s^{-1}$ is the difference in velocity between the rest frame wavelengths of the \Lya\ and \SiIII\ transitions, $\Delta\lambda = 9.17\,$\angstrom.   Within this framework the total power spectrum, $P_{\rm tot}(k)$, including both the \Lya\ and \SiIII\ absorption is given by
\begin{equation}
\begin{aligned}
    P_{\rm tot}(k) &\propto \left| \tilde{\delta}_{\rm Ly\alpha}(k) + \tilde{\delta}_{\rm \SiIII}(k)\right|^{2},\\
     &= P_{\rm Ly\alpha}(k) + P_{\rm \SiIII}(k) + P_{\rm Ly\alpha-\SiIII}(k),
\end{aligned}
\label{eqn:Pclm_old}
\end{equation}
where the first term, $P_{\rm Ly\alpha}(k)$, is the \Lya\ power spectrum, the second term,
\begin{equation} P_{\rm SiIII}(k) = a^{2}P_{\rm Ly\alpha}(k),  \label{eq:PSiIII_old} \end{equation} is the \SiIII\ power spectrum, and the third term,
\begin{equation} P_{\rm Ly\alpha-\SiIII}(k) =2a\cos(\Delta v k)P_{\rm Ly\alpha}(k), \label{eq:Pcross_old} \end{equation}
represents the cross term between Ly$\alpha$ and \SiIII\ that gives the oscillatory behaviour in the total power spectrum.  In the existing literature the normalisation $a$ is often given a redshift dependence $a(z)=f/(1-\bar{F}(z))$, where $\bar F(z)$ is the mean \Lya\ forest transmission, and $f$ is treated as the free parameter.  

We may compare this approach to the \SiIII\ absorption predicted by our hydrodynamical simulation.  Figure~\ref{fig:Flux_Si3compare} shows the directly simulated \SiIII\ transmission (red curve, upper panel) and the \SiIII\ transmission obtained by rescaling the \Lya\ optical depths (lower panel).  More specifically, in the lower panel of Figure~\ref{fig:Flux_Si3compare} we assume $\tau_{\rm \SiIII}(v)=A\tau_{\rm Ly\alpha}(v+\Delta v)$, where we select $A$ so that ${\bar F}_{\rm \SiIII}=\langle e^{-\tau_{\rm \SiIII}} \rangle$ is identical to the hydrodynamical simulation result.  Although this does not follow exactly from Eq.~(\ref{eqn:dclm_old}) (which relies on rescaling $\delta_{\rm Ly\alpha}$, rather than $\tau_{\rm Ly\alpha}$ as we do here)\footnote{Strictly speaking, the flux contrast is proportional to the optical depth, $\delta_{\rm Ly\alpha}\propto \tau_{\rm Ly\alpha}$, only when $\tau_{\rm Ly\alpha}$ is small, such that the first order expansion $F=e^{-\tau_{\rm Ly\alpha}}\simeq 1-\tau_{\rm Ly\alpha}$ is valid.}, it serves to visually illustrate how the \SiIII\ line widths, $b_{\rm \SiIII}$, and hence also $\delta_{\rm\SiIII}$ will be too broad if scaled from the \Lya\ absorption.  Note that for a thermally broadened line
\begin{equation} b_{\rm \SiIII} = \left(\frac{2k_{\rm B}T}{m_{\rm Si}}\right)^{1/2} = \left(\frac{m_{\rm H}}{m_{\rm Si}}\right)^{1/2}b_{\rm Ly\alpha} \simeq 0.19 b_{\rm Ly\alpha}. \label{eq:bvals} \end{equation}

\noindent
We may gain further insight by calculating each of the terms in Eq.~(\ref{eqn:Pclm_old}) for the two cases displayed in Figure~\ref{fig:Flux_Si3compare}.  The results are displayed in Figure~\ref{fig:Pclm_compare}, where the left column shows the power spectrum terms obtained on performing a linear rescaling of the \Lya\ optical depths, while the right column shows the direct simulation of the \SiIII\ absorption.   The value for the linear rescaling parameter $A$ used for the results displayed in the left column is chosen so the amplitude of $P_{\rm \SiIII}(k)$ matches the direct simulation on large scales (i.e, at small $k$). 

There are several key points to highlight in Figure~\ref{fig:Pclm_compare}. First, the agreement between $P_{\rm \SiIII}(k)$ in the rescaled and directly simulated models is reasonable on large scales, $k\lesssim 0.04 \rm\,s\,km^{-1}$. However, the ratio $P_{\rm \SiIII}(k)/P_{\rm Ly\alpha}(k)$ strongly increases toward small scales in the direct simulation.  This contrasts with the scale independent behaviour observed for the linearly rescaled model, and suggests that applying Eq.~(\ref{eqn:dclm_old}) to 1D \Lya\ forest power spectrum constraints at wavenumbers larger than $k\sim 10^{-2}\rm\,s\,km^{-1}$ will result in an underestimate of the contribution of $P_{\rm \SiIII}(k)$ to the total small scale power. A second difference is the oscillatory cross term, $P_{\rm Ly\alpha-\SiIII}(k)$.  Although this retains the same oscillation period in both models, in the direct simulation it has an amplitude that declines towards small scales and makes a negligible contribution to the total power spectrum once $P_{\rm \SiIII}(k)>P_{\rm Ly\alpha-\SiIII}(k)$.  In contrast, the amplitude is constant in the rescaled model (i.e., from Eq.~(\ref{eqn:Pclm_old}), we expect the ratio of the amplitudes for $P_{\rm Ly\alpha-\SiIII}(k)$ and $P_{\rm Ly\alpha}(k)$ to be $a/2$).

The physical origin of these differences arise from several effects missed in the ansatz made by Eq.~(\ref{eqn:dclm_old}).  The first -- as already discussed earlier -- is that the \Lya\ and \SiIII\ line profiles are different.  In general, unless the gas is highly turbulent, the \SiIII\ line widths will be narrower than \Lya\ (cf. Eq.~(\ref{eq:bvals})) resulting in a concomitant increase in small scale power in the \SiIII\ power spectrum. Second, the ratio of the coeval \SiIII\ and \Lya\ optical depths is not constant.   If ignoring redshift space effects by assuming the line profiles are Dirac delta functions, the ratio of the \SiIII\ to \Lya\ optical depths is 
\begin{equation} 
\begin{aligned} 
\beta(\nu)&= \frac{\tau_{\rm \SiIII}(v)}{\tau_{\rm Ly\alpha}(v+\Delta v)},\\ &= \frac{f_{\rm \SiIII} \lambda_{\rm \SiIII}}{f_{\rm Ly\alpha}\lambda_{\rm Ly\alpha}} \frac{x_{\rm \SiIII}(\Delta,T)}{x_{\rm HI}(\Delta,T)}10^{\rm [Si/H]+(Si/H)_{\odot}}, \label{eq:epsilon} 
\end{aligned}
\end{equation}
where $f_{\rm \SiIII}\lambda_{\rm SiIII}/f_{\rm Ly\alpha}\lambda_{\rm Ly\alpha}=3.88$ is the constant ratio of the transition oscillator strengths and rest frame wavelengths \citep{Morton2003}.  At any given redshift we therefore expect $\beta(v)$ to have a pixel dependency arising from variations in the silicon abundance, $[\rm Si/H]$, and \SiIII\ fraction, $x_{\rm \SiIII}$,  where the latter will depend on density, $\Delta$, and temperature, $T$, given some UV background model.  This will impact on the shape and amplitude of both $P_{\rm \SiIII}(k)$ and $P_{\rm Ly\alpha-\SiIII}(k)$. Thirdly, Eq.~(\ref{eqn:dclm_old}) assumes the \Lya\ and \SiIII\ optical depths coherently trace the same underlying structures.  While this is true for the \SiIII\ and \Lya\ optical depths in our models, this does not hold for the observable, $F=e^{-\tau}$.  This is because in practice $\beta(v)\ll 1$ in Eq.~(\ref{eq:epsilon}); intergalactic \SiIII\ absorption lines with $\tau_{\rm \SiIII}\sim 1$ typically arise from much higher density gas compared to \Lya\ lines with $\tau_{\rm Ly\alpha}\sim 1$.  Therefore, even relatively weak \SiIII\ absorbers in our model are coeval with \Lya\ absorption lines that are saturated.

\begin{figure}
    \centering
    \includegraphics[width=\columnwidth]{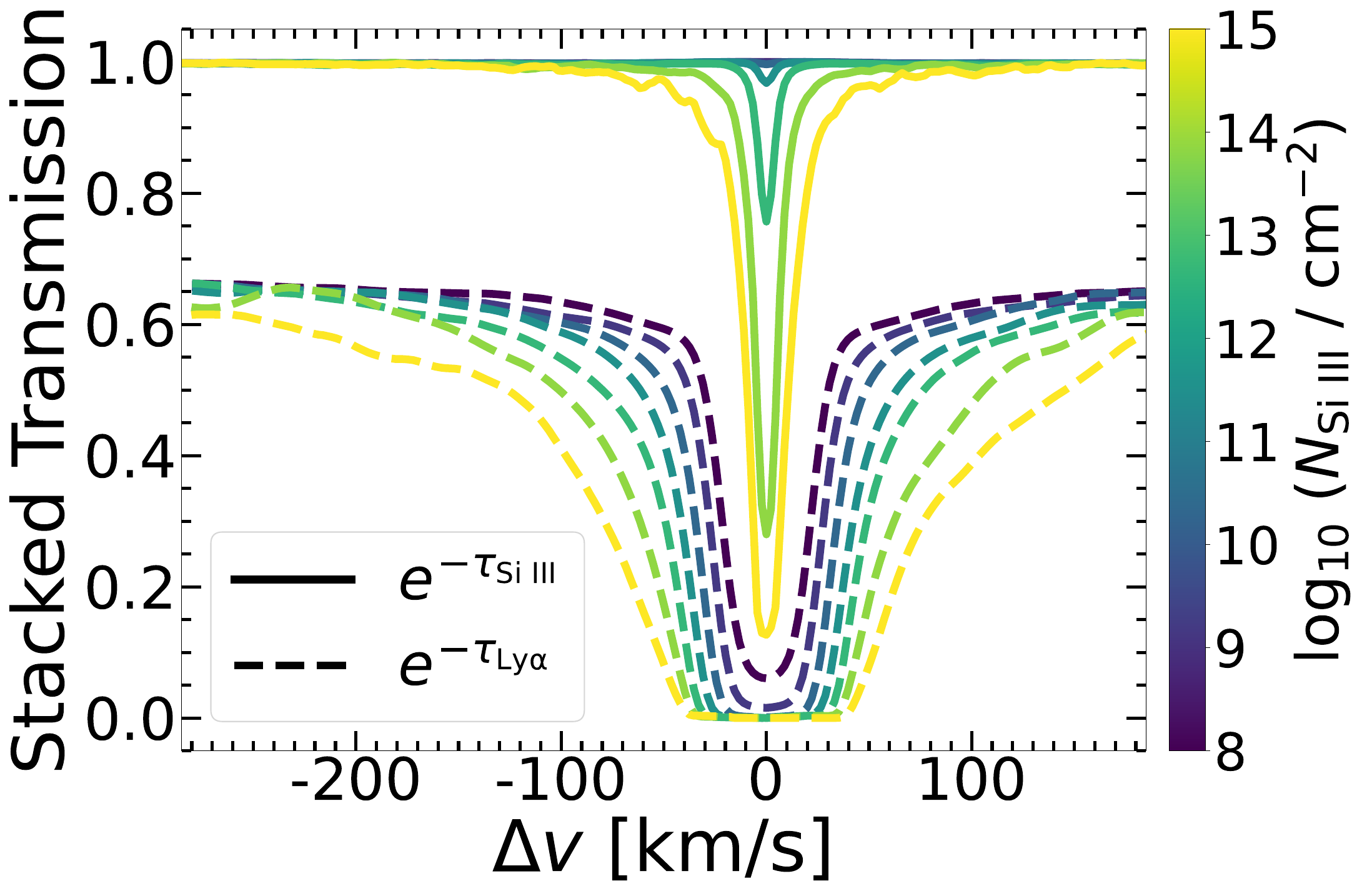}
    \vspace{-0.5cm}
    \caption{Stacked absorption line profiles for different \SiIII\ absorbers as a function of column density, $N_{\rm \SiIII}$, at $z=3.0$.  Solid curves represent the stacked \SiIII\ absorption profiles, while dashed curves show the coeval \Lya\ absorption profiles.  The line colours correspond the $N_{\rm \SiIII}$ in each stack.  For most strong \SiIII\ systems, the \Lya absorption is saturated.  This decorrelates the two tracers and leads to a suppression of the oscillation amplitude in the \Lya--\SiIII\ cross power spectrum toward smaller scales.}
    \label{fig:Decorrelation}
\end{figure}

Figure~\ref{fig:Decorrelation} illustrates this behaviour by comparing coeval \SiIII\ and \Lya\ absorption profiles in different \SiIII\ column density bins, $N_{\rm \SiIII}$, in our simulated spectra.  Here we have estimated $N_{\rm SiIII}$ using the apparent optical depth method \citep{SavageSembach1991}.  For \SiIII\ systems with $\log_{10}(N_{\rm \SiIII}/\rm cm^{-2})\gtrsim 10$, the corresponding \Lya\ forest absorbers are saturated.  This saturation-induced decorrelation between the two tracers manifests as a suppression of the oscillatory amplitude in the cross-power spectrum toward smaller scales, as shown in Figure~\ref{fig:Pclm_compare}.

\subsection{Toward a more accurate model for \SiIII\ contamination in the 1D Ly$\alpha$ forest power spectrum at small scales} 

It is instructive to consider a simple model that approximately captures the behaviour in Figure \ref{fig:Pclm_compare}.  This must take into account two factors: (i) the different line profiles for Ly$\alpha$ and \SiIII\ absorption and (ii) the fact that the ratio between the Ly$\alpha$ and \SiIII\ optical depths will in general depend on temperature and gas density, making the scaling factor $a$ in Eq.~(\ref{eqn:dclm_old}) pixel dependent, instead of a constant.  To incorporate these effects, we modify the original ansatz in Eq.~(\ref{eqn:dclm_old}) to: 

\begin{equation}
    \delta_{\rm \SiIII}(v)\ \circledast \phi(v) = [a \bar{\beta} (1+ \delta_{\beta}(v)) \delta_{\rm Ly\alpha}(v+\Delta v)]. 
\label{eqn:dclm_new}
\end{equation}
Here, $\circledast \phi(v)$ corresponds to a Gaussian kernel convolution that describes the line profile, where for some line broadening scale $b$,
\begin{equation}
\phi(v) = \frac{1}{\sqrt{\pi}\ b}\text{exp}\left(-\frac{v^2}{b^2}\right).
\end{equation}
The perturbative term $\delta_{\beta}(v)$ represents the pixel dependency implied by Eq.~(\ref{eq:epsilon}), where 
\begin{equation} \delta_{\beta}(v)\simeq \frac{\beta(v)}{\bar{\beta}}-1,\end{equation}
if the optical depth is small.  Next, using the standard result
\begin{equation} \tilde\phi(k) = \text{exp}\left(-\frac{b^2 k^2}{4}\right), \end{equation}
the Fourier transform of $\delta_{\rm \SiIII}(v)$ can be obtained by applying a Fourier transform to both sides of Eq.~(\ref{eqn:dclm_new}) and rearranging, yielding
\begin{equation}
\tilde{\delta}_{\rm \SiIII}(k) = a\bar{\beta}\text{exp}\left(\frac{b^{2}k^{2}}{4}+ik\Delta v\right) [\tilde{\delta}_{\rm Ly\alpha}(k) + \tilde{\delta}_{\epsilon}(k)].
\end{equation}
For notational convenience we have written the convolution arising from the Fourier transform of $\delta_{\beta}(v)\delta_{\rm Ly\alpha}(v)$ as $\tilde\delta_{\epsilon}(k)\equiv\tilde{\delta}_{\beta}(k)\circledast \tilde{\delta}_{\rm Ly\alpha}(k)$. The 1D \Lya\ forest power spectrum once again follows from Eq.~(\ref{eqn:Pclm_old}), but the \SiIII\ power spectrum and cross power spectrum now have some additional terms.  For the \SiIII\ power spectrum, we obtain:
\begin{equation} P_{\rm \SiIII}(k) = a^{2}\bar{\beta}^{2}\exp\left(\frac{b^{2}k^{2}}{2}\right)[P_{\rm Ly\alpha}(k) + P_{\rm Ly\alpha-\epsilon}(k) + P_{\rm \epsilon}(k)], \label{eq:PSiIII_new}
\end{equation}
where $P_{\rm Ly\alpha-\epsilon}(k)$ is the power spectrum of the cross-correlation between $\delta_{\rm Ly\alpha}(v)$ and $\delta_{\beta}(v)\delta_{\rm Ly\alpha}(v)$, and $P_{\epsilon}(k)$ is the power spectrum of $\delta_{\beta}(v)\delta_{\rm Ly\alpha}(v)$. The oscillatory cross term is instead given by
\begin{equation} 
P_{\rm Ly\alpha-\SiIII}(k) = 2a\bar{\beta}\cos(\Delta v k)\exp\left(\frac{b^{2}k^{2}}{4}\right)[P_{\rm Ly\alpha}(k)+P_{\rm Ly\alpha-\epsilon}(k)], \label{eq:Pcross_new}
\end{equation} 
where we have used the identity $\cos(\Delta v k)=(e^{i\Delta v k} + e^{-i\Delta v k})/2$.  Comparing Eq.~(\ref{eq:PSiIII_new})--(\ref{eq:Pcross_new}) to Eq.~(\ref{eq:PSiIII_old})--(\ref{eq:Pcross_old}), we therefore expect (i) an additional growing exponential factor that arises from the narrower \SiIII\ line widths,\footnote{Note that we assume constant $b$ in this illustrative model.  In practice, the temperature will vary spatially such that $b$ is also pixel dependent. There may also be a turbulent contribution to the \SiIII\ line widths, which would increase the line broadening scale.}  and (ii) some extra, scale dependent terms that originate from the spatial variations in the silicon abundance and \SiIII\ fraction relative to hydrogen (i.e., Eq.~(\ref{eq:epsilon})).

Overall, these results suggest that an improved model for \SiIII\ contamination in the 1D \Lya\ forest power spectrum at small scales can be provided by a function with four free parameters, 
\begin{equation}
\begin{aligned}
&P_{\rm tot}(k) = P_{\rm Ly\alpha}(k) + P_{\rm \SiIII}(k) + P_{\rm Ly\alpha-\SiIII}(k), \\
    &P_{\rm \SiIII}(k) = a_{\rm auto} \exp\left(\frac{k}{k_{\rm auto}}\right) P_{\rm Ly\alpha}(k), \\
    &P_{\rm Ly\alpha-\SiIII}(k) = a_{\rm cross}\ \text{cos} (\Delta v  k) \exp\left(\frac{k}{k_{\rm cross}}\right)P_{\rm Ly\alpha}(k),
\end{aligned}
\label{eq:model}
\end{equation}
where $a_{\rm auto}$, $a_{\rm cross}$, $k_{\rm auto}$ and $k_{\rm cross}$ are the free parameters. The exponential terms capture the scale dependence of the line broadening and the spatial variations in the silicon abundance and \SiIII\ fraction, and also allow for decorrelation of the \SiIII\ and \Lya\ absorption in the cross power spectrum oscillations (we will show later that $k_{\rm cross}$ is always negative in our models). In the next section we shall demonstrate Eq.~(\ref{eq:model}) provides good approximation to our numerical simulations.


\section{Validation of the improved model} \label{sec:validation}
\subsection{Fit to the fiducial hydrodynamical simulation}

We now test Eq.~(\ref{eq:model}) by performing a direct comparison to the power spectrum computed from the \Lya\ and \SiIII\ absorption in the Sherwood-Relics hydrodynamical simulation described in Section~\ref{sec:method}.

\begin{figure*}
    \includegraphics[width=2\columnwidth]{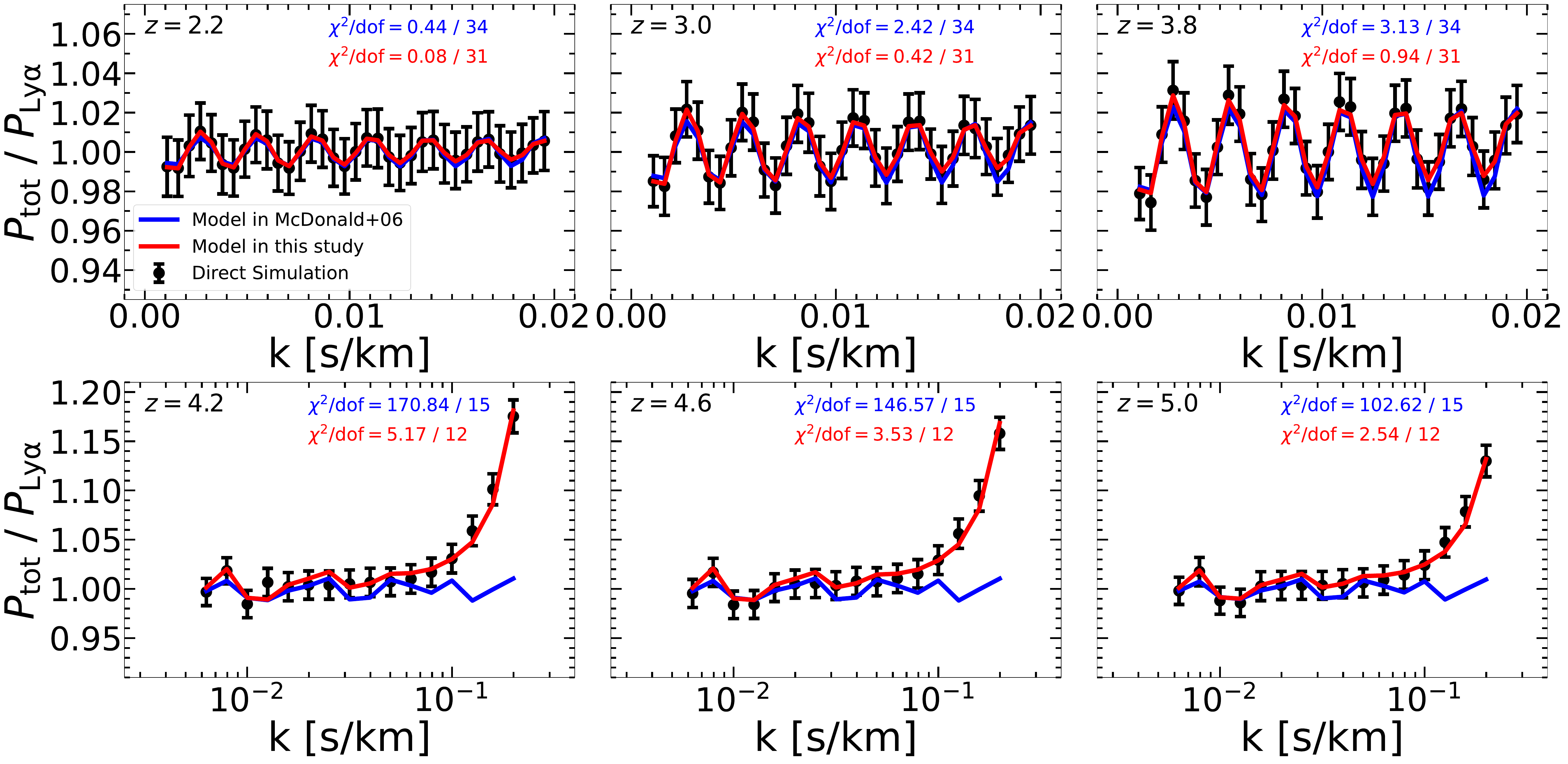}
    \vspace{-0.2cm}
    \caption{Comparison of analytical fits for $P_{\rm tot}(k)/P_{\rm Ly\alpha}(k)$ to the results from a hydrodynamical simulation drawn from the Sherwood-Relics suite (black circles with error bars).  Each panel shows a different redshift, spanning the range $2.2\leq z \leq 5.0$.   The binning with wavenumber in the top row follows the eBOSS 1D power spectrum \citep[][linear binning]{Chabanier2019}, while the bottom row follows the 1D power spectrum measurement from \citet[][logarithmic binning]{Boera2019} obtained from high resolution Keck/HIRES and VLT/UVES spectra.  The error bars give the 68 per cent confidence interval and represent the variance in the 5000 sight-lines of length $40\,h^{-1}\rm\,cMpc$ used to construct the \Lya\ and \SiIII\ absorption spectra.  The red and blue curves respectively show the best fit using the function proposed in this work, Eq.~(\ref{eq:model}), and the original M06 ansatz.   At large scales, $k\lesssim 0.02\rm\,s\,km^{-1}$, both models provide an excellent fit to the simulation.  However, at smaller scales Eq.~(\ref{eq:model}) performs much better because the contribution from the \SiIII\ power spectrum, $P_{\rm \SiIII}(k)$, becomes significant.}
    \label{fig:Fits_diff_surveys}
\end{figure*}

Our first task is to assess how our proposed model fares relative to the approach used in the existing literature, given by Eq.~(\ref{eqn:Pclm_old})--(\ref{eq:Pcross_old}) \citep{Mcdonald2006}.  The results of this test are summarised in Figure \ref{fig:Fits_diff_surveys}, where we compare our model to simulation data (black circles with error bars) binned to match the scales probed by 1D power spectrum measurements from eBOSS \citep[][top row]{Chabanier2019} and high resolution data from Keck/HIRES and VLT/UVES \citep[][bottom row]{Boera2019}.  We consider six different redshift bins spanning $2.2\leq z \leq 5.0$.  The error bars on $P_{\rm tot}(k)/P_{\rm Ly\alpha}(k)$ show the 68 per cent confidence interval obtained by bootstrapping the simulated sight-lines.  Note these error bars represent the variance in the simulation and do not correspond to expected uncertainties on observational data (for the latter, see Section~\ref{sec:comparedata}). 

The top three panels in Figure~\ref{fig:Fits_diff_surveys} use the eBOSS binning. This only extends to $k \sim 0.02\rm\,s\,km^{-1}$ due to the low resolution and signal-to-noise of these data.  The blue and red curves show the best fits for Eq.~(\ref{eqn:Pclm_old})--(\ref{eq:Pcross_old}) M06 and our proposed modification in Eq.~({\ref{eq:model}), respectively.  The minimum $\chi^{2}$ and degrees of freedom are displayed in the inset; the fits are extremely good.  The $\chi^{2}$ values for the model proposed here are slightly smaller than those of M06. This is because our model reproduces the modest decline in the oscillation amplitude, which the M06 model fails to capture. Hence, at the scales typically probed by low to moderate resolution \Lya\ forest spectra, the small-scale enhancement to power expected from the \SiIII\ absorption is negligible.  Reassuringly, this also confirms that the M06 ansatz remains a reasonable approximation for modelling the \SiIII\ oscillations in the power spectrum at $k \lesssim 0.02\rm\,s\,km^{-1}$.  Furthermore, the best fit to our hydrodynamical simulation requires $a=0.008$ at $z=3$ which is consistent with the value inferred from eBOSS, $a=0.018\pm 0.012$ \citep{Chabanier2019}.\footnote{We compute this from $f_{\rm \SiIII}=0.006\pm0.004$ reported by \citet{Chabanier2019}, where they impose a redshift dependence $a=f_{\rm \SiIII}/[1-{\bar F}(z)]$ in their parameterisation.  Here we have assumed the mean transmission from \citet{Becker2013} at $z=3$, given by ${\bar F}=0.669$.}

\begin{table*}
\centering
\begin{tabular}{cccccc} 
\hline
z & $k_{\rm auto}/10^{-2} \rm\,s\,km^{-1}$ & $k_{\rm cross}/10^{-2} \rm\,s\,km^{-1}$ & $a_{\rm auto}/10^{-3}$ & $a_{\rm cross}/10^{-2}$ & $\langle|r(k)|\rangle$ \\ [0.5ex] 
\hline\hline
2.2 & $2.96$ & $-1.52$ & $0.99$ & $1.30$ & 0.0063 \\[1ex]
2.4 & $3.06$ & $-1.52$ & $1.14$ & $1.46$ & 0.0049 \\[1ex]
2.6 & $3.22$ & $-1.47$ & $1.49$ & $1.91$ & 0.0050 \\[1ex]
2.8 & $3.48$ & $-1.59$ & $2.13$ & $2.36$ & 0.0059 \\[1ex]
3.0 & $3.84$ & $-1.58$ & $3.27$ & $2.79$ & 0.0058 \\[1ex]
3.2 & $4.23$ & $-1.77$ & $4.13$ & $3.33$ & 0.0060 \\[1ex]
3.4 & $4.62$ & $-1.88$ & $4.91$ & $3.66$ & 0.0063 \\[1ex]
3.6 & $4.99$ & $-2.00$ & $5.61$ & $3.83$ & 0.0067 \\[1ex]
3.8 & $5.22$ & $-2.14$ & $5.65$ & $3.64$ & 0.0062 \\[1ex]
4.0 & $5.30$ & $-2.25$ & $5.44$ & $3.74$ & 0.0057 \\[1ex]
4.2 & $5.63$ & $-2.25$ & $5.46$ & $3.60$ & 0.0058 \\[1ex]
4.4 & $5.81$ & $-2.42$ & $5.91$ & $3.42$ & 0.0061 \\[1ex]
4.6 & $5.85$ & $-2.32$ & $5.58$ & $3.31$ & 0.0059 \\[1ex]
4.8 & $5.90$ & $-2.36$ & $5.30$ & $3.17$ & 0.0054 \\[1ex]
5.0 & $5.91$ & $-2.33$ & $4.80$ & $3.12$ & 0.0055 \\[1ex]
\hline
\end{tabular}
\caption{Best-fit values of Eq. (\ref{eq:model}) to the 1D power spectrum predicted by our fiducial hydrodynamical simulation data at redshifts $2.2\leq z \leq 5.0$. The parameters $k_{\rm auto}$ and $a_{\rm auto}$ are fit simultaneously to $P_{\rm \SiIII}(k)/P_{\rm Ly\alpha}(k)$ within the range $k < 0.2\rm\,s\,km^{-1}$, while $k_{\rm cross}$ and $a_{\rm cross}$ are fit simultaneously to $P_{\rm Ly\alpha-\SiIII}(k)/P_{\rm Ly\alpha}(k)$ within the range $k < 0.06\rm\,s\,km^{-1}$ (see text for details). The last column reports the absolute value of the fractional deviation between the simulation and the fit for wavenumbers $k < 0.2\rm\,s\,km^{-1}$.}
\label{tab:fitting}
\end{table*}

However, when including smaller scales using the high-resolution binning in the bottom three panels of Figure~\ref{fig:Fits_diff_surveys}, the enhancement in small-scale power arising from the \SiIII\ absorption becomes relevant.   In this case Eq.~(\ref{eq:model}) continues to provide a very good approximation to $P_{\rm tot}(k)/P_{\rm Ly\alpha}(k)$, but the M06 model fails to capture this behaviour and the fit to the simulation becomes increasingly poor toward smaller scales / larger wavenumbers.   Our revised model, given by Eq.~(\ref{eq:model}), thus provides a more accurate template for modelling the correlated \SiIII\ absorption.

The best fit parameters in Eq.~(\ref{eq:model}) to the power spectrum from our fiducial simulation are summarised Table \ref{tab:fitting}. Here, $k_{\rm auto}$ and $a_{\rm auto}$ are obtained by fitting $P_{\rm \SiIII}(k)/P_{\rm Ly\alpha}(k)$ at $k < 0.2\rm\,s\,km^{-1}$.  For $k_{\rm cross}$ and $a_{\rm cross}$, however, we restrict the fit to $P_{\rm Ly\alpha-\SiIII}(k)/P_{\rm Ly\alpha}(k)$ to smaller wavenumbers, $k < 0.06 \rm \,s\,km^{-1}$.  This upper limit is chosen to ensure the fit to $P_{\rm Ly\alpha-\SiIII}(k)$ is sensitive to the oscillation amplitude in the cross term, rather than noise at small scales where we expect $P_{\rm \SiIII}(k)>P_{\rm Ly\alpha-\SiIII}(k)$ (see Figure~\ref{fig:Pclm_compare}).  At each redshift, we also report the absolute value of the fractional deviation, $\langle |r(k)| \rangle)$, between the simulation and best fit, where
\begin{equation} 
r(k) = \frac{\left[P_{\rm tot}(k)/P_{\rm Ly\alpha}(k)\right]_{\rm sim}}{\left[P_{\rm tot}(k)/P_{\rm Ly\alpha}(k)\right]_{\rm fit}} - 1.  
\label{eq:r(k)}
\end{equation}
For $k<0.2\rm\,s\,km^{-1}$, $\langle |r(k)| \rangle$ is always within 1 per cent of $P_{\rm tot}(k)/P_{\rm Ly\alpha}(k)$ obtained from the hydrodynamical simulation.

\subsection{The effect of varying silicon abundance and temperature} \label{sec:TandZ}

Our second task is to test the generality of Eq.~(\ref{eq:model}); how well does our revised analytical fit perform when confronted with simulations constructed using different silicon abundances and temperatures (i.e., the parameters that primarily set the column densities and line widths of the \SiIII\ absorption lines)?  We test this by contrasting our fiducial model with spectra constructed assuming a different IGM silicon abundance, $\rm [Si/H]$, and/or IGM temperatures that have been rescaled in post-processing following the approach used by  \citet{Irsic2024}.   

It is straightforward to change the $\rm [Si/H]$ used in our fiducial model.  We will consider four alternative values for $\rm [Si/H]$ here.  The first two bracket the $\rm [Si/H]$ constraints from \citet{Schaye2003,Aguirre2004} as given by Eq.~(\ref{eq:SiH}) and correspond to the 1$\sigma$ lower and upper bounds on that measurement.  Note, however, this does require the  uncertain extrapolation of this measurement from $z=4.1$ to $z=5$. The other two models adopt constant values\footnote{We show in Appendix~\ref{app:zscale_test} that our fiducial silicon abundance model gives very similar results to a model with constant $\rm[Si/H]=-3.0$.} of $\rm [Si/H]=-4.0$ and $-3.5$.  The temperature scaling is instead performed by rotating and translating the temperature-density plane in our fiducial simulation\footnote{Recall that our fiducial thermal history is provided by the \citet{Puchwein2019} UV background model (see their figure 7).} around the power-law temperature-density relation, $T=T_{0}\Delta^{\gamma-1}$, which holds at $\Delta\leq 10$ \citep{HuiGnedin1997, McQuinnUptonSanderbeck2016}.  We vary this by $\Delta\log_{10}T_{0}=\pm 0.2$ and $\Delta\gamma=\pm0.2$, giving a range similar to the $1\sigma$ uncertainties on measurements of the IGM thermal state at $2< z< 6$ \citep{Gaikwad2020, Gaikwad2021}. By cross-varying $\rm [Si/H]$, $T_{0}$ and $\gamma$, we therefore generate $25$ sets of simulated spectra per redshift bin. These variations are explored for 15 redshift bins, ranging from $z=2.2$ to $5.0$ in steps of $0.2$, yielding a total of $375$ simulated datasets encompassing different redshifts, silicon abundances and gas temperatures. 

We fit Eq.~(\ref{eq:model}) to each dataset within the range $k < 0.2\rm\,s\,km^{-1}$.  The behaviour of the best-fit parameters at $z=3$ is summarised in Figure~\ref{fig:Fits_diff_data}. 
The parameter $k_{\rm cross}$ is always negative, consistent with a \Lya\--\SiIII\ oscillation amplitude that declines toward smaller scales.  It is mainly sensitive to variation in the thermal parameter $T_{0}$, with more rapid damping of the oscillations (less negative $k_{\rm cross}$) for hotter gas.  This happens because the coeval \Lya\ absorbers -- which are typically on the flat part of the curve of growth (see Figure~\ref{fig:Decorrelation}) -- are broadened as the gas temperature increases.  This has the effect of shifting the decorrelation between \SiIII\ and \Lya\ to larger scales.   The parameter $k_{\rm auto}$ is instead primarily sensitive to the variation in $T_{0}$, and is only weakly dependent on the silicon abundance, $\rm [Si/H]$, and the slope of the temperature density relation, $\gamma$.  Hotter temperatures increase the thermal broadening of the \SiIII\ and \Lya\ absorption features, yielding a smaller $k_{\rm auto}$ (or equivalently, a larger smoothing scale $\sim 1/k_{\rm auto}$).  Note this produces an \emph{enhancement} in $P_{\rm tot}(k)/P_{\rm Ly\alpha}(k)$ at small scales. Colder temperatures have the opposite effect, suppressing the additional small scale power from the \SiIII\ absorbers in $P_{\rm tot}(k)/P_{\rm Ly\alpha}(k)$.\footnote{Another way to think of this is that hotter (colder) temperatures suppress (enhance) small scale power for both $P_{\rm Ly\alpha}(k)$ and $P_{\rm \SiIII}(k)$, but \emph{$P_{\rm Ly\alpha}(k)$ is always suppressed (enhanced) to a greater degree} than $P_{\rm \SiIII}$ because of the mass dependence of the thermal broadening kernel (see Eq.~(\ref{eq:bvals})).}   The insensitivity of $P_{\rm tot}(k)/P_{\rm Ly\alpha}(k)$ to $\gamma$ is likely because most of the \Lya\ forest arises from gas close to the mean density, $\Delta=1$ \citep[see e.g., table 3 in][]{Becker2011}.  

In contrast, variations in silicon abundance primarily affect the amplitude parameters $a_{\rm auto}$ and $a_{\rm cross}$, while leaving $k_{\rm auto}$ and $k_{\rm cross}$ largely unchanged.  The silicon abundance $\rm [Si/H]$ directly rescales the \SiIII\ optical depths and hence the \SiIII\ power spectrum.  A higher silicon abundance increases the \SiIII\ contribution to the total absorption, thereby raising both the \SiIII\ power spectrum and $P_{\rm tot}(k)/P_{\rm Ly\alpha}(k)$.  The apparent correlation between $a_{\rm auto}$ and $a_{\rm cross}$ in the lower left panel of Figure~\ref{fig:Fits_diff_data} also motivated us to explore the possibility of using a simple linear scaling to reduce these two amplitudes to a single parameter.  Unfortunately, however, on careful testing we found this linear scaling does not remain fixed on varying the wavenumber range used for the fit.  A four parameter model is thus required for generality.

\begin{figure}
    \includegraphics[width=0.48\textwidth]{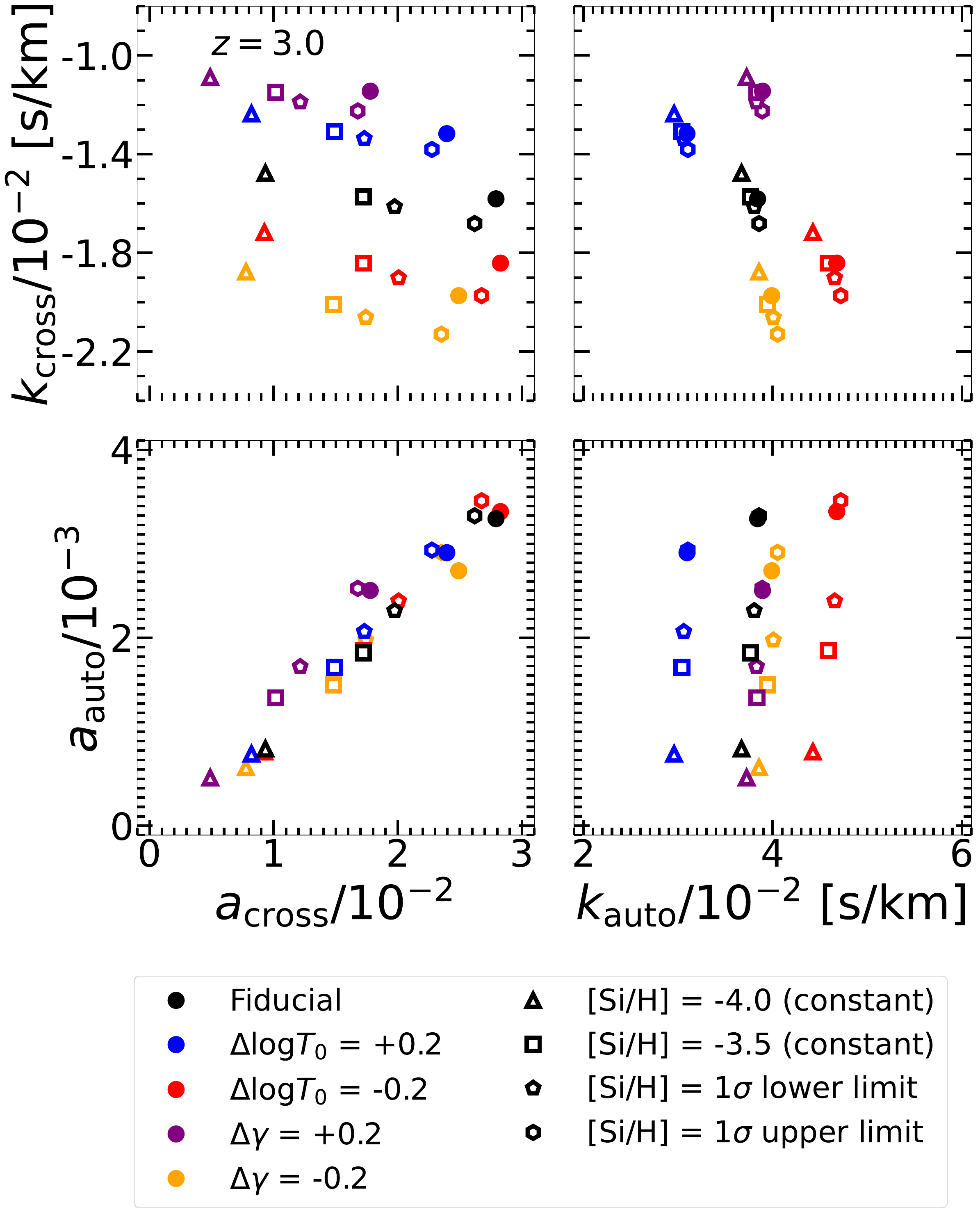}
    \vspace{-0.5cm}
    \caption{The best-fit values of $k_{\rm auto}$, $k_{\rm cross}$, $a_{\rm auto}$, and $a_{\rm cross}$ (see Eq.~(\ref{eq:model})) at $z=3$.  The models are constructed using different silicon abundances and gas temperatures (see text for details). The temperature variations are indicated by different colours, while the silicon abundance variations are indicated by the different marker styles.  Changing the temperature at mean density, $T_{0}$, primarily affects $k_{\rm auto}$ and $k_{\rm cross}$, while changing the silicon abundance affects $a_{\rm auto}$ and $a_{\rm cross}$.}
    \label{fig:Fits_diff_data}
\end{figure}

Over the full redshift range considered here, $2.2\leq z \leq 5.0$, the absolute value of the fractional deviation, $\langle |r(k)| \rangle$, for the best fit parameters in Eq.~(\ref{eq:model}) to each model are displayed in Figure~\ref{fig:Fits_allz}.  In general, the fit quality degrades for higher temperatures and larger silicon abundances, most likely because the assumption of $\sim \exp(k/k_{\rm auto})$ scaling made in Section~\ref{sec:modelling} begins to break down.   The fit quality is also poorer at lower redshift as the \Lya\ forest becomes more transmissive.   Nevertheless, $\langle |r(k)| \rangle$ remains below 2 per cent for all models (and below 1 per cent for the fiducial model), indicating that Eq.~(\ref{eq:model}) continues to provide a very good description of the simulation data across all tested conditions.

\begin{figure}
    \includegraphics[width=0.48\textwidth]{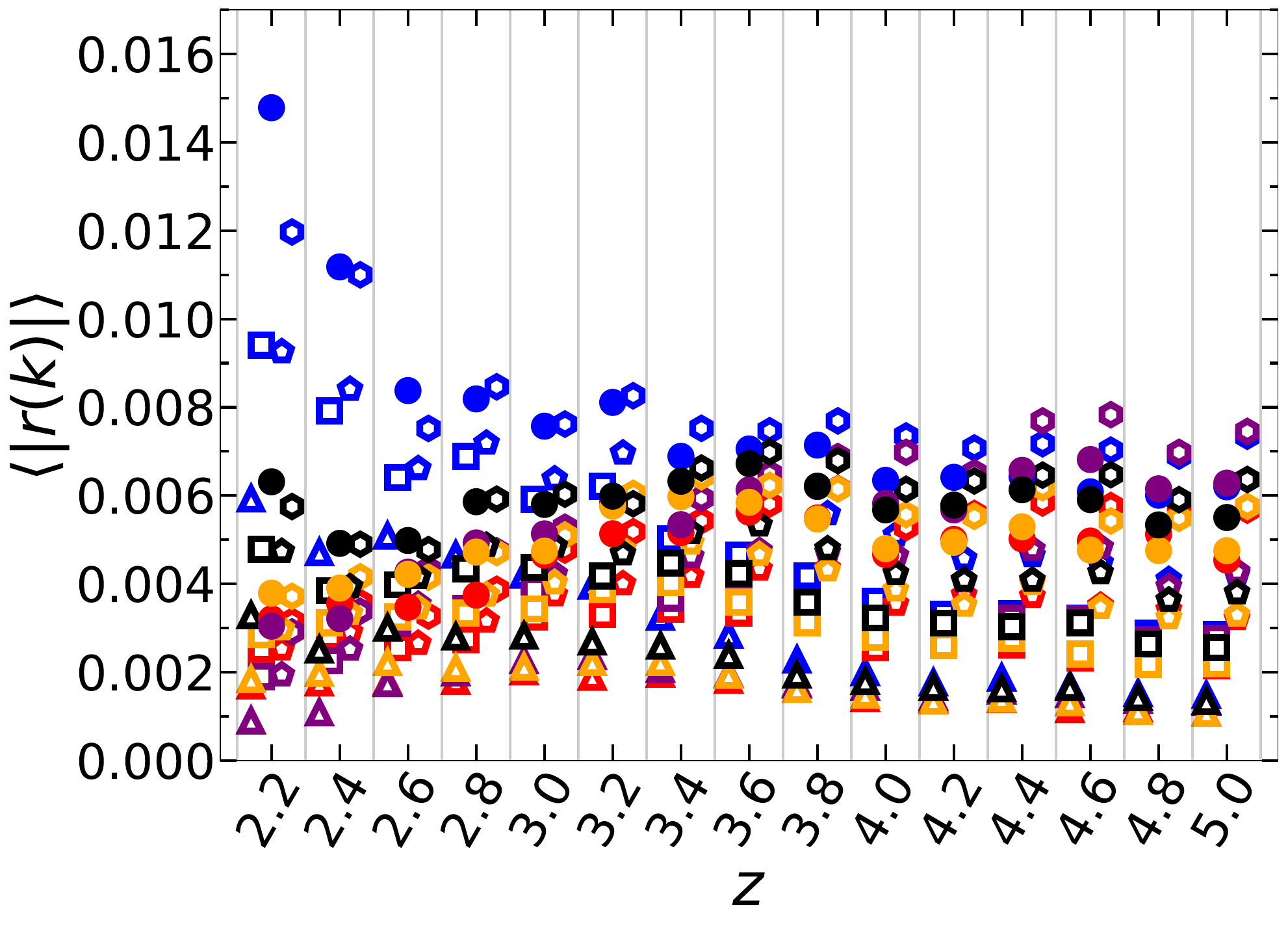}
    \vspace{-0.6cm}
    \caption{The absolute value of the fractional deviation, $\langle |r(k)| \rangle$, obtained when fitting Eq.~(\ref{eq:model}) to simulation data at $k<0.2\rm\,s\,km^{-1}$ constructed assuming different silicon abundances and temperatures as a function of redshift (see text for details). The colour and marker styles are the same as those listed in the legend of Figure~\ref{fig:Fits_diff_data}.}
    \label{fig:Fits_allz}
\end{figure}

\begin{figure*}
    \includegraphics[width=0.9\textwidth]{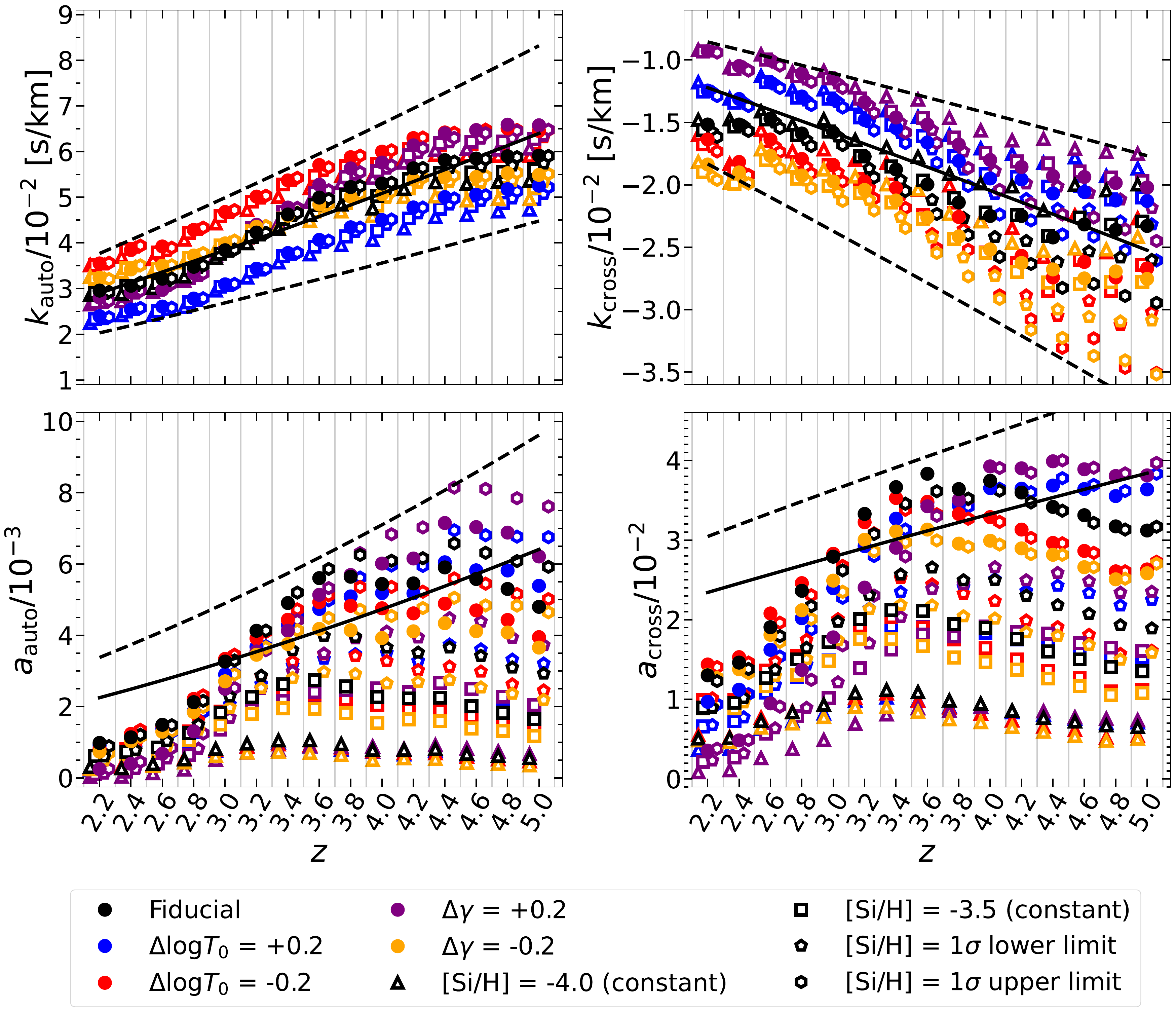}
    \vspace{-0.3cm}
    \caption{The redshift evolution of the best-fit parameters for Eq.~(\ref{eq:model}) to numerical simulations of $P_{\rm tot}(k)/P_{\rm Ly\alpha}(k)$ at $k<0.2\rm\,s\,km^{-1}$ that assume different silicon abundances and IGM temperatures. The colour and marker styles follow those in Figure~\ref{fig:Fits_allz}.  The solid black lines are power-law approximations to the redshift evolution of the fiducial model, where $k_{\rm auto}(z)=3.84\times 10^{-2}[(1+z)/4]^{1.26}\rm\,s\,km^{-1}$, $k_{\rm cross}(z)=-1.58\times 10^{-2}[(1+z)/4]^{1.15}\rm\,s\,km^{-1}$, $a_{\rm auto}(z)=3.27\times 10^{-3}[(1+z)/4]^{1.66}$, and $a_{\rm cross}(z)=2.79\times 10^{-2}[(1+z)/4]^{0.79}$.  The dashed grey lines show these power laws scaled by a constant to encompass the various models and thus give a rough indication of the prior range, where $0.7 \leq A \leq 1.3$  for $Ak_{\rm auto}(k)$, $0.7 \leq B \leq 1.5$  for $Bk_{\rm cross}(k)$, $0 \leq C \leq 1.5$ for $Ca_{\rm auto}(k)$, and $0 \leq D \leq 1.3$ for $Da_{\rm cross}(k)$}.
    \label{fig:Fits_zEvol}
\end{figure*}

\subsection{The effect of varying redshift} \label{sec:redvary}

Finally, in Figure~\ref{fig:Fits_zEvol} we examine how the best fit parameters in Eq.~(\ref{eq:model}) vary with redshift.  We first consider $k_{\rm auto}$ in the upper left panel.  For all models we find $k_{\rm auto}$ increases toward higher redshift.  This behaviour is driven by the redshift evolution of the IGM temperature and pixel-to-pixel variations in the \SiIII\ absorption.  Recall from Section~\ref{sec:method} that $k_{\rm auto}$ characterises the scale dependence of $P_{\rm \SiIII}(k)$, which has contributions from both the pixel-dependent optical depth ratio (i.e. Eq.~(\ref{eq:epsilon})) and thermal broadening. The former depends on the redshift evolution of the silicon abundance, $\rm [Si/H]$, and the \SiIII\ fraction, $x_{\rm \SiIII}$. The latter introduces a redshift dependence according to $k_{\rm auto}(z) \propto b^{-2}(z) \propto T^{-1}(z)$, or equivalently $k_{\rm auto}(z) \propto T_0^{-1}(z)$ if the $\gamma$ redshift dependence is weak.  In our fiducial model the pixel dependency has a redshift evolution that approximately follows the behaviour of $a_{\rm auto}(z)$ in the lower left panel of Figure~\ref{fig:Fits_zEvol} (see discussion below) whereas $T_0(z)$ peaks at $z \sim 3$ due to photo-heating from \HeII\ reionisation \citep[see e.g.][]{Puchwein2019,Gaikwad2021}. In contrast, we find $k_{\rm cross}$ decreases toward higher redshift so that the cross term oscillations are damped less rapidly at earlier times.  In our model this is driven by the increasing abundance of \SiIII\ toward higher redshift.  This shifts the coeval \Lya\ absorption lines to lower column densities and pushes the decorrelation between \SiIII\ and \Lya\ to smaller scales.

Next, we consider the amplitude parameters, $a_{\rm auto}$ and $a_{\rm cross}$.  Following the pioneering work by M06, previous studies have typically imposed a redshift dependence $a(z)=f/[1-\bar{F}(z)]$ where $\bar{F}(z)$ is the mean transmission in the \Lya\ forest and $f$ is a free parameter.   Under this assumption the amplitude of the \SiIII\ power spectrum declines toward higher redshift as $\bar{F}$ decreases.  However, in this work we find the opposite trend in the simulation data; the amplitude increases with redshift at $z\lesssim 3.5$ and remains relatively flat at higher redshifts, as shown by the redshift evolution of $a_{\rm auto}$ and $a_{\rm cross}$ in the lower panels of Figure~\ref{fig:Fits_zEvol}.   We have further checked this by fitting both our model and the original M06 ansatz to our simulations at $k < 0.02\rm\,s\,km^{-1}$. The redshift evolution for the amplitude parameters (not shown here) yielded the same trend, confirming this is not simply a consequence of including small scale power in the fit.  Note again the changes in $a_{\rm auto}$ and $a_{\rm cross}$ are insensitive to thermal variations and their redshift evolution is driven almost entirely by the silicon abundance, $\rm [Si/H]$, and/or the \SiIII\ fraction, $x_{\rm \SiIII}$. 

Overall, this implies the redshift evolution of $a_{\rm auto}$ and $a_{\rm cross}$ is primarily driven by $x_{\rm \SiIII}$ in our models.  This is corroborated by the redshift evolution of $x_{\rm \SiIII}$ shown in the right panel of Figure~\ref{fig:ionfrac}, which exhibits a qualitatively similar trend to $a_{\rm auto}$ and $a_{\rm cross}$.   It is therefore challenging for the empirically calibrated UV background synthesis model used in this work \citep{Puchwein2019} to produce a \SiIII\ power spectrum and cross-term amplitude that scales as $[1-\bar{F}(z)]^{-1}$ if the silicon abundance does not evolve strongly with redshift.  Counteracting the $x_{\rm \SiIII}$ evolution would require $\rm [Si/H]$ to increase at least as rapidly as the corresponding decrease in $x_{\rm \SiIII}$ toward lower redshift as the UV background spectrum hardens (see Appendix~\ref{app:zscale_test}).

It is, nevertheless, useful to parameterise the redshift evolution of $k_{\rm auto}$, $k_{\rm cross}$, $a_{\rm auto}$ and $a_{\rm cross}$ to facilitate marginalisation over multiple redshift bins.  One possibility is to adopt the redshift scaling of our fiducial model in Table~\ref{tab:fitting}.  This is model-dependent, however, and may be overly restrictive.  A more flexible but slightly less accurate alternative is to approximate the redshift evolution using power laws \citep[see e.g.][for a similar approach]{Viel2013}.  We use $k_{\rm auto}(z)=3.84\times 10^{-2}[(1+z)/4]^{1.26}\rm\,s\,km^{-1}$, $k_{\rm cross}(z)=-1.58\times 10^{-2}[(1+z)/4]^{1.15}\rm\,s\,km^{-1}$, $a_{\rm auto}(z)=3.27\times 10^{-3}[(1+z)/4]^{1.66}$, and $a_{\rm cross}(z)=2.79\times 10^{-2}[(1+z)/4]^{0.79}$ to approximate our fiducial model. These are shown as the black solid lines in Figure~\ref{fig:Fits_zEvol}. The dashed black lines are scaled by a constant to encompass the various models and thus give a rough indication of the prior range for the amplitude of these relations, where $0.7 \leq A \leq 1.3$  for $Ak_{\rm auto}(k)$, $0.7 \leq B \leq 1.5$  for $Bk_{\rm cross}(k)$, $0 \leq C \leq 1.5$ for $Ca_{\rm auto}(k)$, and $0 \leq D \leq 1.3$ for $Da_{\rm cross}(k)$.  The amplitude priors are derived from the $1\sigma$ uncertainties on the observational constraints on the gas temperature and silicon abundance in the IGM (see Section~\ref{sec:TandZ}).  Note also that we adopt only upper limits for the priors on $a_{\rm auto}(z)$ and $a_{\rm cross}(z)$ to account for the possibility the power spectrum is consistent with no \SiIII\ absorption.}  The slopes of these power-laws may also be varied if gas temperature and silicon abundance redshift evolutions that differ from our fiducial model are required.  Note, however, that both the redshift evolution of the UV background (which impacts the \SiIII\ fraction, $x_{\rm \SiIII}$) and IGM temperature at $2\leq z \leq 5$ are both relatively well constrained by observation \citep{Puchwein2019,Gaikwad2021}.

\subsection{Recommended implementation} \label{sec:usage}

\begin{figure*}
    \centering
    \includegraphics[width=0.85\textwidth]{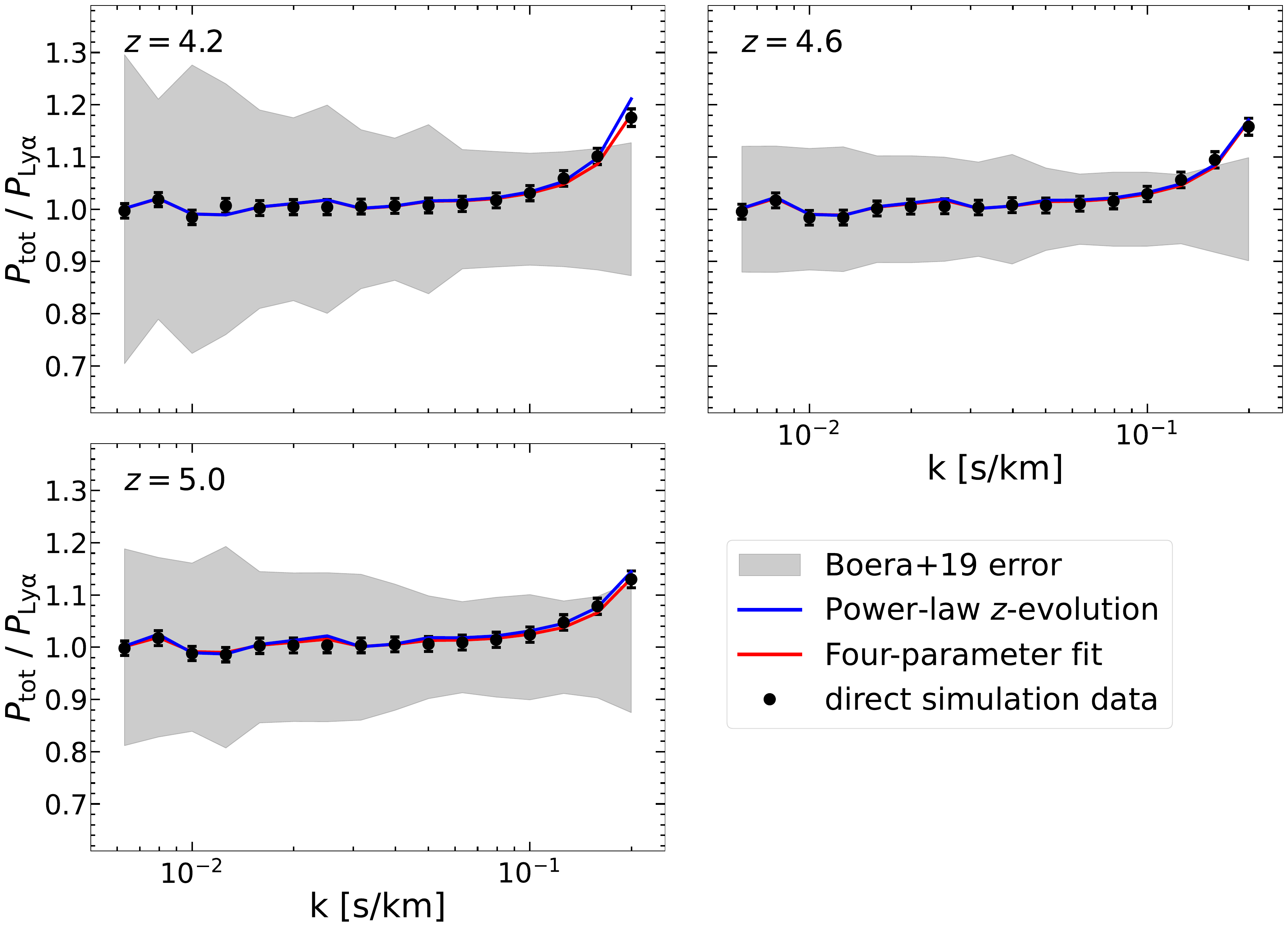}
    \vspace{-0.3cm}
    \caption{Comparison of our fiducial model with the fractional uncertainties, $\sigma/P_{\rm Ly\alpha}$, on the 1D power spectrum measurements from \citet[][grey shaded regions]{Boera2019}. The red curves represent direct four-parameter fit to Eq.~(\ref{eq:model}) ({\bf Method I}), while the blue curves represent fits imposing power-law redshift dependencies ({\bf Method II}) described in Section~\ref{sec:usage}. We also overplot the direct simulation results using black dots. The error bars represent the 68 per cent confidence interval on the simulation data, matching the results in the bottom panels of Figure~\ref{fig:Fits_diff_surveys}.  The small-scale power enhancement due to \SiIII\ absorption exceeds the \citet{Boera2019} uncertainty in the largest $k$-bins ($k = 0.2 \rm \,s\,km^{-1}$) for all three redshift bins.} 
    \label{fig:ObsErrs}
\end{figure*}
Here we summarise the implementation of the \SiIII\ model presented in this work.  The full correction to the \Lya\ only power spectrum, $P_{\rm Ly\alpha}(k)$, as a function of wavenumber, for $k\leq 0.2\rm\,s\,km^{-1}$, and redshift, for $2.2\leq z\leq 5.0$, is given by
 
\begin{equation} 
\begin{aligned} 
\frac{P_{\rm tot}(k,z)}{P_{\rm Ly\alpha}(k,z)} &=  1 + Ca_{\rm auto}(z)\exp \left(\frac{k}{Ak_{\rm auto}(z)}\right) \\
 &+ Da_{\rm cross}(z)\exp \left(\frac{k}{Bk_{\rm cross}(z)}\right)\cos(\Delta v k) , \label{eq:finalfit}
\end{aligned}
\end{equation}
where $A$, $B$, $C$, and $D$ scale the redshift dependent parameters $k_{\rm auto}(z)$, $k_{\rm cross}$, $a_{\rm auto}(z)$, and $a_{\rm cross}(z)$, and $\Delta v = 2271\rm\,km\,s^{-1}$ is the velocity separation of coeval \Lya\ and \SiIII\ absorption.  We suggest reasonable prior ranges on $A$, $B$, $C$, and $D$ are $0.7 \leq A \leq 1.3$, $0.7 \leq B \leq 1.5$, $0 \leq C \leq 1.5$, and $0 \leq D \leq 1.3$ (where $C,D=0$ corresponds to no \SiIII\ contribution to the power) based on the range of models considered in Figure~\ref{fig:Fits_allz}.  Note, however, these prior ranges may be extended if a larger range of IGM temperatures or silicon abundances are of interest.  There are then two choices for the redshift-dependent parameters:

\begin{enumerate}

    \item \textbf{Method I}: Our preferred approach is to adopt the redshift scaling for $k_{\rm auto}(z)$, $k_{\rm cross}$, $a_{\rm auto}(z)$, and $a_{\rm cross}(z)$ given by the interpolation of the data in Table~\ref{tab:fitting}.  There are four free parameters in Eq.~(\ref{eq:finalfit}): $A$, $B$, $C$, and $D$.  For our fiducial IGM model $A=1$, $B=1$, $C=1$, and $D=1$.  The advantage of this approach is that it better captures the redshift evolution in our fiducial IGM model but is less flexible than an approximate power-law redshift scaling.\\
    
    \item \textbf{Method II}: Adopt the approximate power-law redshift scalings $k_{\rm auto}(z)=3.84\times 10^{-2}[(1+z)/4]^{1.26}\rm\,s\,km^{-1}$, $k_{\rm cross}(z)=-1.58\times 10^{-2}[(1+z)/4]^{1.15}\rm\,s\,km^{-1}$, $a_{\rm auto}(z)=3.27\times 10^{-3}[(1+z)/4]^{1.66}$, and $a_{\rm cross}(z)=2.79\times 10^{-2}[(1+z)/4]^{0.79}$, again giving four free parameters in Eq.~(\ref{eq:finalfit}): $A$, $B$, $C$, and $D$. For additional flexibility the slopes of the power-laws may also be varied to allow for IGM temperatures and silicon abundances that have redshift dependencies that differ significantly from our fiducial model. Note, however, that both the IGM temperature and UV background at $2.2\leq z \leq 5$ are observationally well-constrained and these parameters should therefore be varied judiciously (see also Appendix~\ref{app:zscale_test}).\\

\end{enumerate}

\section{Implications for warm dark matter constraints} \label{sec:comparedata}

\begin{figure*}
    \centering
    \includegraphics[width=0.95\textwidth]{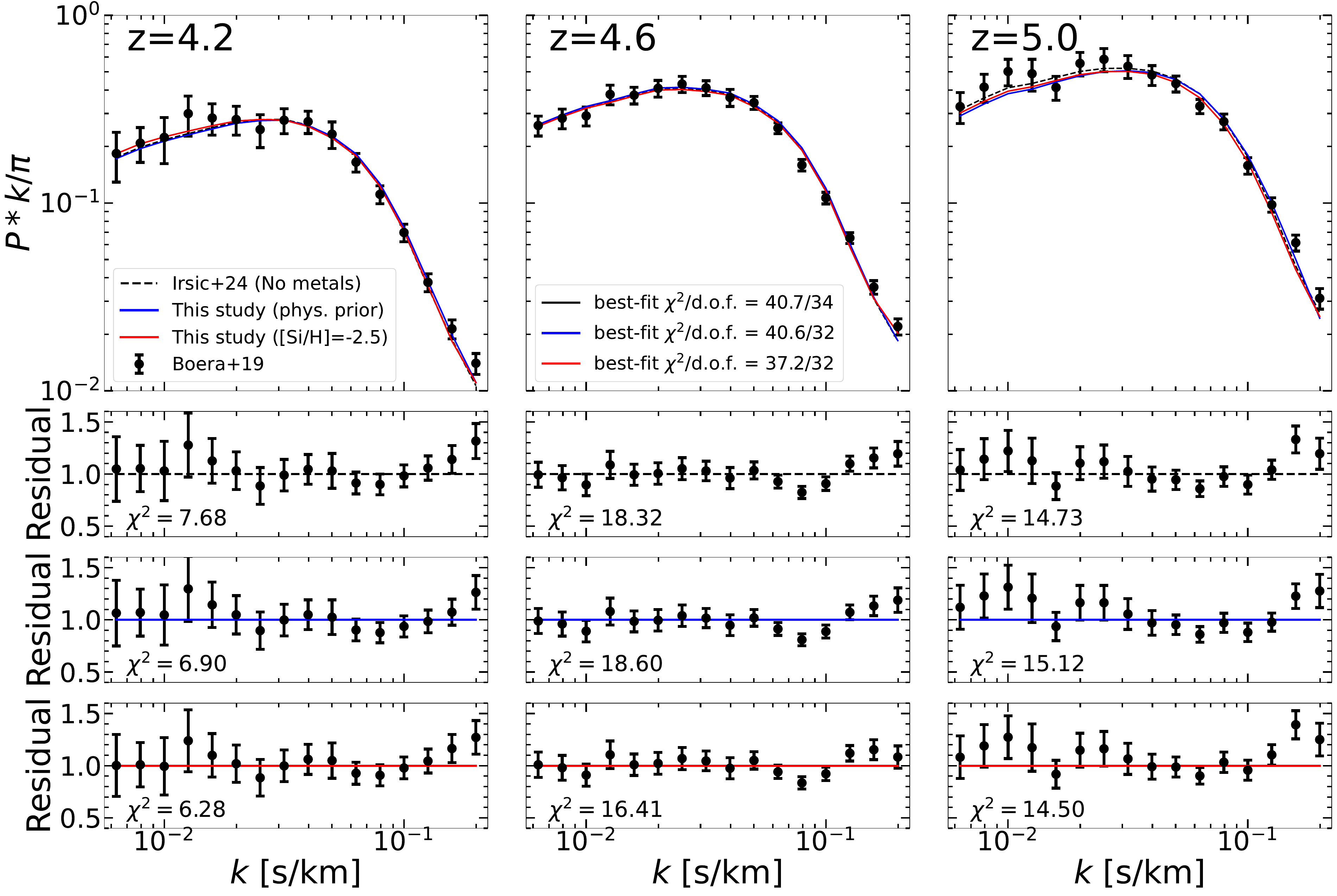}
    \vspace{-0.3cm}
    \caption{Comparison of the \citet{Irsic2024} best-fit model (black dashed curves) to our \SiIII\ model using two different priors (red and blue solid curves). The blue curves show best-fit model using the physical priors described in Section~\ref{sec:TandZ}, while the red curves use a broader prior range, equivalent to allowing a larger silicon abundance of [Si/H]=$-2.5$.  The black data points are the 1D power spectrum measurements from \citet{Boera2019} in three redshift bins. The lower panels show the residuals of the \citet{Boera2019} measurements with respect to the models. }
    \label{fig:WDM}
\end{figure*}

To assess the implications of our correlated \SiIII\ absorption model for existing warm dark matter constraints from the 1D Ly$\alpha$ forest power spectrum,  we compare our model to the uncertainties on the \citet{Boera2019} power spectrum measurement (grey shading) in Figure~\ref{fig:ObsErrs}.  The contribution from \SiIII\ becomes increasingly prominent at small scales. The red and blue curves show, respectively, our best-fit fiducial models using \textbf{Method I} and \textbf{Method II} described in Section~\ref{sec:usage}.  The black circles show the results from our hydrodynamical simulation with the same error bars displayed in Figure~\ref{fig:Fits_diff_surveys}.  At scales $k \sim 0.2\rm\,s\,km^{-1}$, the small-scale power from the \SiIII\ power spectrum -- which was not accounted for in earlier models -- becomes comparable to the observational error bars.

It is thus interesting to investigate the impact of our model on existing warm dark matter (WDM) constraints from 1D \Lya\ power spectrum measurements. \cite{Irsic2024} recently used the measurement of \cite{Boera2019} to set a stringent lower bound on the mass of a warm dark matter (WDM) thermal relic. Their best-fit WDM model showed excellent agreement at intermediate scales ($k < 0.1$ s km$^{-1}$; $\chi^2$/dof = 20.4/20), giving a limit of $m_{\rm WDM} \geq 4.1$ keV (at 95 per cent confidence).  However, extending their analysis to smaller scales ($k < 0.2$ s km$^{-1}$) resulted in a somewhat poorer fit ($\chi^2$/dof = 40.7/34) with a limit of $m_{\rm WDM} \geq 5.7$ keV. \cite{Irsic2024} noted the larger reduced $\chi^2$ was due to additional small-scale power at $k > 0.1$ s km$^{-1}$  -- precisely where our study reveals substantial \SiIII\ driven enhancements to the small-scale power.

To quantify the contribution of correlated \SiIII\ absorption identified in this study, we incorporate our model (using {\bf Method I}) into the \cite{Irsic2024} pipeline and compare the best-fit models for the 1D \Lya\ power spectrum with and without our \SiIII\ model.  We present this comparison in Figure~\ref{fig:WDM}.  We find that $k_{\rm auto}$ and $a_{\rm auto}$ are prior-dominated, while $k_{\rm cross}$ and $a_{\rm cross}$ remain unconstrained. This is due to the dominance of the small-scale enhancement from $P_{\rm \SiIII}(k)$ over the cross-term oscillations for the wavenumbers considered here.  When using physically motivated priors as described in Section~\ref{sec:usage}, the $\chi^2$/dof changes from $40.7/34$ in \citet{Irsic2024} to $40.6/32$, shown by the blue curves in Figure~\ref{fig:WDM}.\footnote{Note that because $k_{\rm cross}$ and $a_{\rm cross}$ are unimportant for the fit in this example, the degrees of freedom are effectively 32 instead of 30, despite now having four extra free parameters.} Consequently, the goodness-of-fit remains similar and the constraint on the WDM mass slightly weakens from $m_{\rm WDM} \geq 5.71$ keV to $m_{\rm WDM} \geq 5.65$ keV.  To better fit the small-scale enhancement, we also try a looser prior range on the silicon abundance such that $0 \leq C \leq 3$ and $0 \leq D \leq 2$, equivalent to allowing $[\rm Si/H]\sim -2.5$ (this is within $3 \sigma$ of our fiducial silicon abundance model).  The results are shown by the red curves in Figure~\ref{fig:WDM}. In this case, the $\chi^2$/dof is $37.2/32$, and the constraint on the WDM mass now weakens from $m_{\rm WDM} \geq 5.71$ keV to $m_{\rm WDM} \geq 5.58$ keV.

In summary, for physically motivated priors our \SiIII\ model slightly reduces the $\chi^{2}$ for the \citet{Boera2019} power spectrum on small scales and does not significantly change the existing constraints on warm dark matter from \citet{Irsic2024}. Note, however, the goodness-of-fit remains similar due to the additional free parameters.  If instead adopting a prior equivalent to $\rm [Si/H]\sim -2.5$, the goodness-of-fit improves slightly and the \citet{Irsic2024} warm dark matter constraints weakens by $0.07$ keV.  The silicon abundance is observationally uncertain, however, and it is unclear whether this level of IGM metal enrichment is warranted at $z\sim 5$.  Regardless, our results suggests that the small-scale power enhancement observed in the \citet{Boera2019} data by \citet{Irsic2024} may at least be partially attributable to correlated \SiIII\ absorption.   We also expect this effect will become more relevant in the future as the precision of constraints on small-scale power at $z>4$ improve.


\section{Summary} \label{sec:summary}

We have used a hydrodynamical simulation drawn from the Sherwood-Relics simulation suite \citep{Bolton2017,Puchwein2023} to construct an improved model for the impact of correlated \SiIII\ absorption on the 1D \Lya\ forest power spectrum.  We have identified a significant enhancement in the 1D power spectrum due to \SiIII\ at small scales ($k \gtrsim 0.06\rm\,s\,km^{-1}$ s/km) that is not explicitly accounted for in previous work. To address this limitation, we have developed a physically motivated fitting function to extend the canonical framework for modelling the \Lya-\SiIII\ correlation first introduced by M06.  Our model takes into account two critical effects missed previously -- the fact that \SiIII\ line profiles are narrower than those of Ly$\alpha$, and that the ratio of coeval Ly$\alpha$ and \SiIII\ optical depths are not constant. 

Our improved model yields a four parameter fit (see Eq.~(\ref{eq:finalfit})) where $k_{\rm auto}$ controls the small-scale power enhancement, $k_{\rm cross}$ controls the damping of the \Lya--\SiIII\ cross power spectrum oscillation amplitude, and $a_{\rm auto}$ and $a_{\rm cross}$ independently control the amplitudes of the \SiIII\ power spectrum and the \Lya\--\SiIII\ cross-correlation power spectrum, respectively.  We validate this model by comparison to multiple simulated datasets at redshifts $2.2\leq z \leq 5.0$ for wavenumbers $k\lesssim 0.2\rm\,s\,km^{-1}$.  The absolute value of the fractional deviation between our fitting function and all tested datasets is within 2 per cent, suggesting good agreement between our fitting function and the simulation data across a variety of IGM thermal and metal enrichment histories.   Incorporating this effect will be essential for future cosmological and astrophysical constraints using high resolution \Lya\ forest data.  We provide full guidelines for the implementation of the model within existing parameter inference frameworks in Section~\ref{sec:usage}.

We also find the redshift dependence reported in M06 typically assumed for the amplitude of the \SiIII\ and \Lya--\SiIII\ cross power spectra -- such that $a(z) \propto [1-\bar{F}(z)]^{-1}$ where $\bar{F}(z)$ is the mean transmission in the \Lya\ forest and $f$ is a free parameter -- is not observed in our fiducial model.   Under this assumption the amplitude of the \SiIII\ oscillations declines toward higher redshift as $\bar{F}$ decreases.  Instead, our empirically constrained simulations return the opposite trend; the amplitude increases with redshift at $2\leq z\leq 3.5$ and remains approximately constant at higher redshifts.   This is primarily due to the hardening of our assumed model for the metagalactic UV background spectrum at $z\lesssim 3.5$ for photon energies around the \SiIII\ ionisation potential of $33.5\rm\,eV$, coupled with modest evolution in the intergalactic silicon abundance at $2<z<5$ \citep{Schaye2003,Aguirre2004}.  The origin of this spectral hardening is the increasingly important contribution from quasar emission to the UV background \citep[e.g][]{Khaire2019,Puchwein2019}.   Counteracting this trend would require an intergalactic silicon abundance that increases more rapidly toward lower redshift than has been assumed in our fiducial model, and/or a UV background model with a softer ionising spectrum.

Finally, our \SiIII\ absorption model has little impact on existing constraints on warm dark matter free streaming if adopting physically motivated priors on the IGM temperature and silicon abundance, but it may partially account for the otherwise difficult-to-fit small scale power enhancement at $k>0.1\rm\,s\,km^{-1}$ in the \citet{Boera2019} power spectrum measurements noted by \citet{Irsic2024}.  We anticipate that modelling this effect will become more important as the precision of the observational data improves.  Our results also present an opportunity to place novel constraints on the metal enrichment of the low density IGM using the 1D power spectrum.  We intend to explore this possibility in future work.

\section*{Acknowledgements}

KM gratefully acknowledges the financial support provided by the China Scholarship Council programme (Project ID: CSC NO. 202309110040).  The hydrodynamical simulation used in this work was performed using the Joliot Curie
supercomputer at the Tr{\'e}s Grand Centre de Calcul (TGCC) and the
Cambridge Service for Data Driven Discovery (CSD3), part of which is
operated by the University of Cambridge Research Computing on behalf
of the STFC DiRAC HPC Facility (www.dirac.ac.uk).  We acknowledge the
Partnership for Advanced Computing in Europe (PRACE) for awarding us
time on Joliot Curie in the 16th call. The DiRAC component of CSD3 was
funded by BEIS capital funding via STFC capital grants ST/P002307/1
and ST/R002452/1 and STFC operations grant ST/R00689X/1.  DiRAC is part of the National e-Infrastructure. JSB is supported by STFC consolidated grant ST/X000982/1.  We thank Volker Springel for making P-Gadget-3 available and Philip Parry for technical support.


\section*{Data Availability}
All data and analysis code used in this work are available from the
first author on reasonable request.  Further guidance on accessing the publicly available Sherwood-Relics simulation data may also be found at
\url{https://www.nottingham.ac.uk/astronomy/sherwood-relics/}.



\bibliographystyle{mnras}
\bibliography{bibliography}

\newpage
\appendix

\section{The redshift evolution of the silicon abundance} \label{app:zscale_test}

\begin{figure*}
    \includegraphics[width=0.9\textwidth]{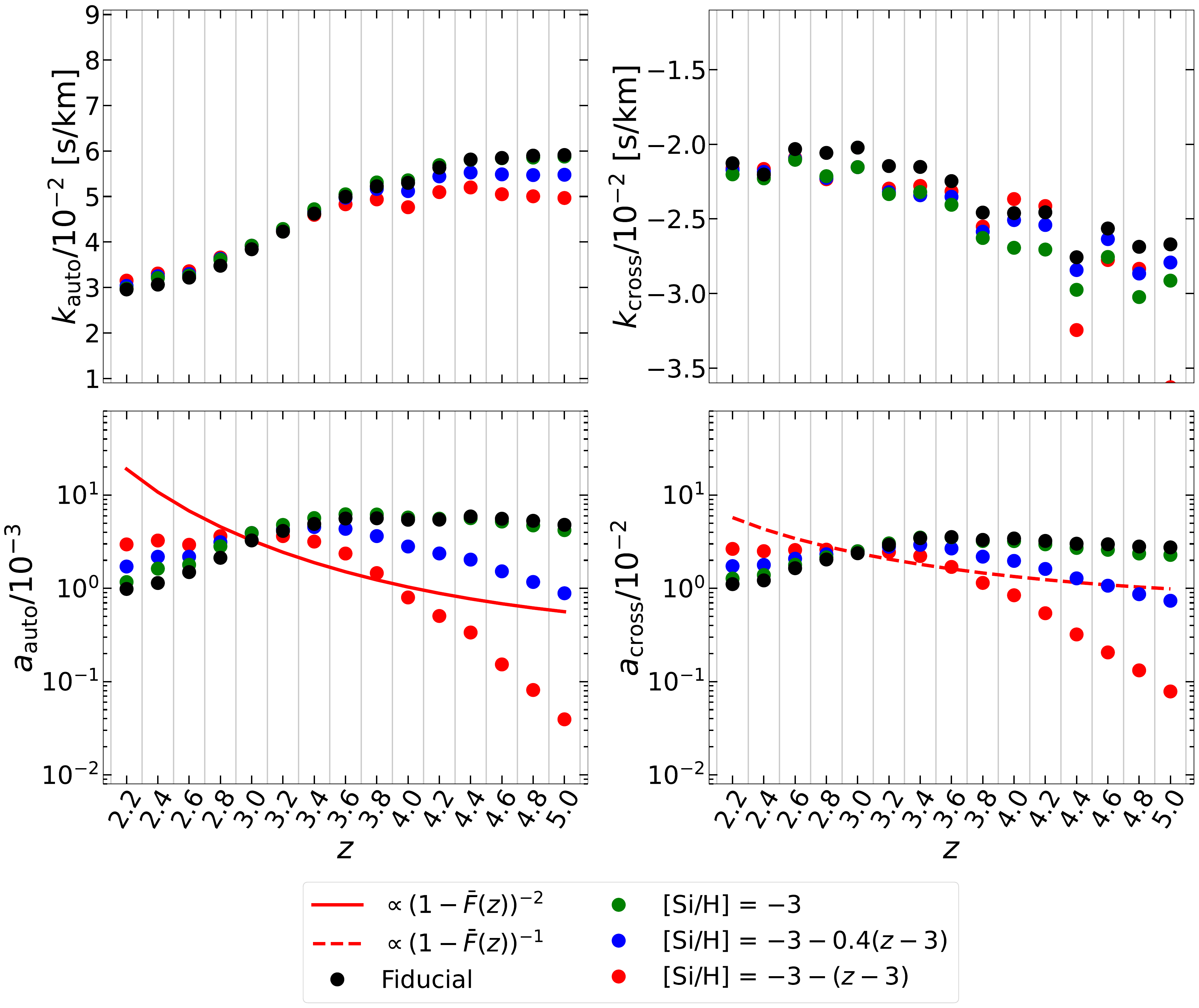}
    \vspace{-0.3cm}
    \caption{The best-fit values of $k_{\rm auto}$, $k_{\rm cross}$, $a_{\rm auto}$, and $a_{\rm cross}$ obtained for four different models for the intergalactic silicon abundance $\rm [Si/H]$.  In all cases the \SiIII\ fraction, $x_{\rm \SiIII}$ is computed using the \citet{Puchwein2019} UV background synthesis model.  The black points denote our fiducial model for the silicon abundance (Eq.~\ref{eq:SiH}) where $\rm [Si/H] \propto 0.08(z-3)$, whereas the green points assume a constant silicon abundance, $\rm [Si/H]=-3$.  The blue and red points instead correspond to a $\rm [Si/H]$ that declines respectively by a factor of $10$ and $10^3$ over $z=2$--$5$.  For comparison the red solid and dashed curves follow the redshift scaling usually assumed in the literature, $ a_{\rm \SiIII} \sim a^{2}(z) \propto [1-\bar{F}(z)]^{-2}$ (lower left panel) and $ a_{\rm Ly\alpha- \SiIII} \sim a(z) \propto [1-\bar{F}(z)]^{-1}$ (lower right panel), where $\bar{F}(z)$ is the mean \Lya\ forest transmission.}
    \label{fig:test_Fits_zEvol}
\end{figure*}

In Section~\ref{sec:redvary} we demonstrated that the redshift evolution of $a_{\rm auto}$ and $a_{\rm cross}$ is largely driven by the \SiIII\ fraction, $x_{\rm \SiIII}$, when assuming a constant intergalactic silicon abundance $\rm [Si/H]$.  Furthermore, our fiducial model for $\rm [Si/H]$  \citep[][see Eq.~(\ref{eq:SiH})]{Schaye2003,Aguirre2004} has a weak redshift dependence.  Hence, in both cases, the hardening of the intergalactic radiation field during \HeII\ reionisation at $z\sim 3$ causes the \SiIII\ power spectrum and cross-term amplitudes to \emph{increase} from $z=2$ to $z\simeq 3.5$ and remain almost constant at $z\gtrsim3.5$.  This contrasts with the usual $[1-\bar{F}(z)]^{-1}$ scaling used in the literature \citep[e.g.][]{Mcdonald2006,Palanque-Delabrouille2013,Chabanier2019}.

We consider this further in Figure~\ref{fig:test_Fits_zEvol}, where we examine the effect of varying the redshift dependence of the silicon abundance, $\rm [Si/H]$, on the redshift evolution of the best-fit parameters in Eq.~\eqref{eq:model}.   We consider our fiducial model (black circles), a constant $\rm [Si/H]=-3$ (green points) and a toy model where the silicon abundance decreases by a factor $10^{3}$ from $z=2$ to $z=5$, $\rm [Si/H]=-3-(z-3)$ (red points).  Note this is an extreme model and is inconsistent with the pixel optical depth constraints.  In contrast, over the same redshift interval the total mass of metals traced by \HI\ absorption systems decreases by a factor of $\sim 10$ in the EAGLE hydrodynamical simulation \citep[][their figure 5]{Rahmati2018}.  A similar factor of $\sim 10$ decrease is observed for gas in the simulations presented by \citet{Shen2010} (their figure 6).  For the final toy model (blue dots) we therefore decrease the silicon abundance by a factor $\sim 10$ from $z=2$ to $z=5$.   

For $k_{\rm auto}$ and $k_{\rm cross}$, shown in the upper panel of Figure~\ref{fig:test_Fits_zEvol}, variations in the $\rm [Si/H]$ redshift evolution have a comparatively minor effect, with the most noticeable impact occurring at higher redshifts ($z > 3$).  In the lower panels, for constant $\rm [Si/H]=-3$ the evolution of $a_{\rm auto}$ and $a_{\rm cross}$ is primarily driven by the \SiIII\ fraction $x_{\rm \SiIII}$, as already discussed in Section~\ref{sec:redvary}.  Consequently, these parameters increase toward higher redshift (green dots), mirroring the redshift dependence of $x_{\rm \SiIII}$ in Figure~\ref{fig:ionfrac}. Our fiducial model (black dots) coincides with the constant $\rm [Si/H]=-3$ case because the redshift dependence in Eq.~\eqref{eq:SiH} is weak ($\rm [Si/H]\propto 0.08(z-3)$). To reverse the redshift evolution of $a_{\rm auto}$ and $a_{\rm cross}$ so that they decrease toward higher redshifts – as required by the canonical scaling, shown by the red solid and dashed curves – requires $\rm [Si/H]$ to decrease toward higher redshift.

\section{The effect of varying the maximum wavenumber} \label{app:kcut_test}

\begin{figure*}
    \includegraphics[width=0.9\textwidth]{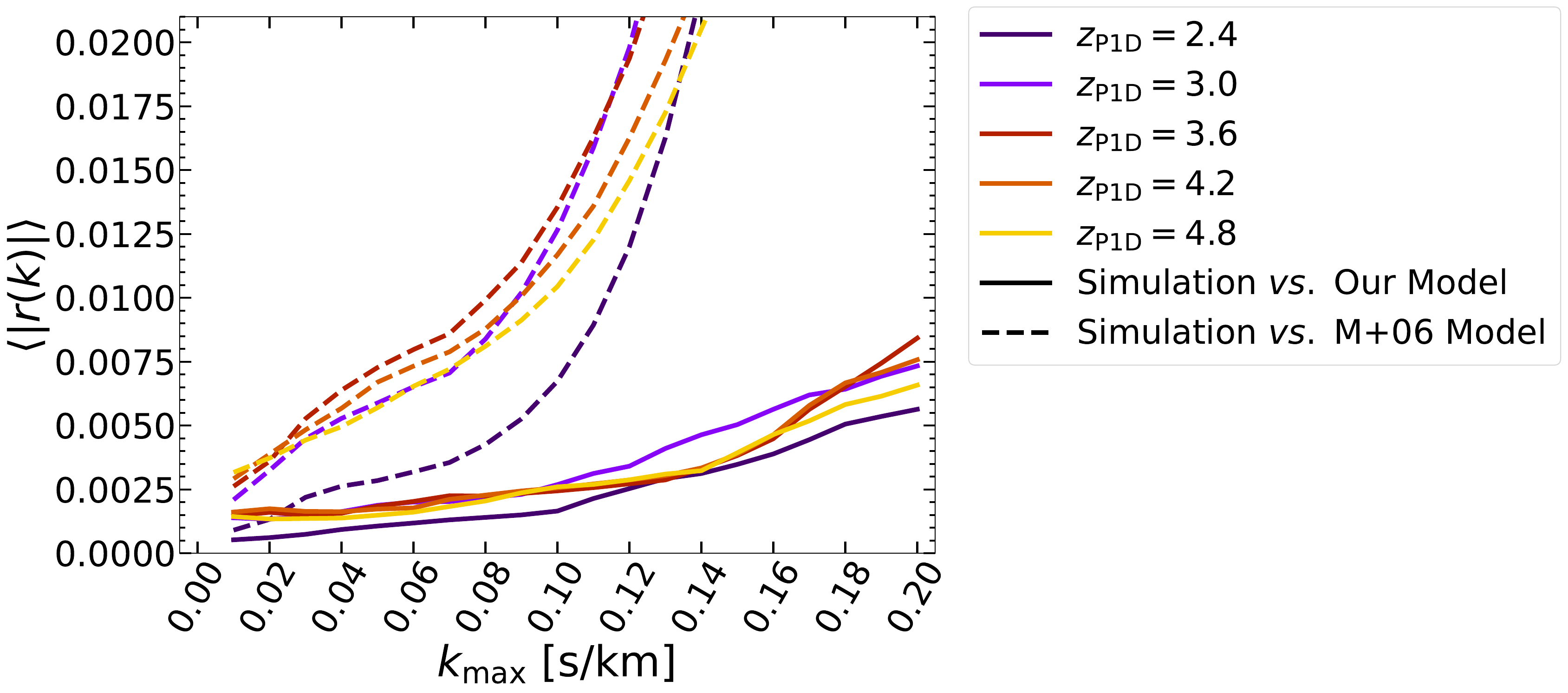}
    \vspace{-0.3cm}
    \caption{The absolute value of the fractional deviation for fits to the 1D power spectrum from our fiducial simulation using either the improved model proposed in this work (solid curves) or the original M06 ansatz (dashed curves).  The results are shown as a function of the maximum wavenumber used in the fit, $k_{\rm max}$.  The different colours represent the redshift bins shown in the legend.  Our model demonstrates good agreement with the simulation within 1 per cent at all redshifts for all $k_{\rm max}$ values.}
    \label{fig:Varykcut}
\end{figure*}

In Section~\ref{sec:validation} we quantified the accuracy of our fit in Figure~\ref{fig:Fits_allz} for wavenumbers $k \leq 0.2\rm\,s\,km^{-1}$. To investigate the impact of varying the maximum wavenumber, $k_{\rm max}$, of the simulated power spectrum we fit to, in Figure~\ref{fig:Varykcut}, we additionally show the absolute value of the fractional deviation for the fit as a function of $k_{\rm max}$ across five different redshift bins. The solid curves correspond to the four-parameter fit used in this work (Eq.~\ref{eq:model}), while the dashed curves are for the M06 ansatz. Our model remains consistent with the simulation within 1 per cent regardless of redshift or $k_{\rm max}$. This demonstrates that our model is applicable across all scales considered and successfully captures both the declining oscillatory behaviour from the cross power spectrum and the small-scale enhancement from the \SiIII\ power spectrum. For comparison, when fitting the M06 model to the simulation data, the relative errors are well below 1 per cent at $k_{\rm max} < 0.06\rm\,s\,km^{-1}$, but increase rapidly above that threshold as the small-scale enhancement due to the \SiIII-\SiIII\ term becomes significant.

\end{document}